\documentclass[acmlarge]{acmart}

\AtBeginDocument{%
  \providecommand\BibTeX{{%
    \normalfont B\kern-0.5em{\scshape i\kern-0.25em b}\kern-0.8em\TeX}}}

\usepackage{lipsum}

\usepackage{longtable}
\usepackage{multirow}
\usepackage{tcolorbox}
\newtcolorbox[auto counter]{mybox}[2][]{%
title=Example~\thetcbcounter: #2, #1}

\begin{document}

\title{Attackers reveal their arsenal: An investigation of adversarial techniques in CTI reports}


\author{Md Rayhanur Rahman}
\email{mrahman@ncsu.edu}
\affiliation{%
  \institution{NC State University}
  \city{Raleigh}
  \state{NC}
  \country{USA}
}

\author{Setu Kumar Basak}
\email{sbasak4@ncsu.edu }
\affiliation{%
  \institution{NC State University}
  \city{Raleigh}
  \state{NC}
  \country{USA}
}

\author{Rezvan Mahdavi Hezaveh}
\email{rezvan.mahdavi@gmail.com}
\affiliation{%
  \institution{NC State University}
  \city{Raleigh}
  \state{NC}
  \country{USA}
}

\author{Laurie Williams}
\email{lawilli3@ncsu.edu}
\affiliation{%
  \institution{NC State University}
  \city{Raleigh}
  \state{NC}
  \country{USA}
}


\begin{abstract}
\textbf{Context:} Cybercrime groups, driven by financial and geopolitical motives, launch advanced persistent threat (APT) attacks. The attacks consist of adversarial techniques, which adversaries perform step-by-step by during cyberattacks. Cybersecurity vendors often publish cyber threat intelligence (CTI) reports, referring to the written artifacts on technical and forensic analysis of the techniques used by the malware in APT attacks. To defend organizations, prevalent techniques used by malware in APT attacks and the association among the techniques need to be identified. \textbf{Objective:} \textit{The goal of this research is to inform cybersecurity practitioners about how adversaries form cyberattacks through an analysis of adversarial techniques documented in cyberthreat intelligence reports.} \textbf{Dataset:} We use 594 adversarial techniques cataloged in MITRE ATT\&CK. We systematically construct a set of 667 CTI reports that MITRE ATT\&CK used as citations in the descriptions of the cataloged adversarial techniques. \textbf{Methodology:} We analyze the frequency and trend of adversarial techniques, followed by a qualitative analysis of the implementation of techniques. Next, we perform association rule mining to identify pairs of techniques recurring in APT attacks. We then perform qualitative analysis to identify the underlying relations among the techniques in the recurring pairs. \textbf{Findings:} The set of 667 CTI reports documents 10,370 techniques in total, and we identify 19 prevalent techniques accounting for 37.3\% of documented techniques. We also identify 425 statistically significant recurring pairs and seven types of relations among the techniques in these pairs. The top three among the seven relationships suggest that techniques used by the malware inter-relate with one another in terms of (a) abusing or affecting the same system assets, (b) executing in sequences, and (c) overlapping in their implementations. We identify that obtaining information on the operating and network system of the victim environment is the most prevalent technique and appears in the highest number of recurring pairs. We identify that spear-phishing is the most prevalent way of initial infection. We also identify three prevalent misuses of system functionalities: macro in office documents, registry in Windows, and task scheduler. We also identify that mimicking legitimate users through compromised credentials is the most prevalent persistence-related technique used by malware. Overall, the study quantifies how adversaries leverage techniques through malware in APT attacks based on publicly reported documents. We advocate organizations prioritize their defense against the identified prevalent techniques and actively hunt for potential malicious intrusion based on the identified pairs of techniques. 
\end{abstract}




\maketitle

\section{Introduction}

Information technology (IT) systems draw continuous attention from cyberattackers with financial motives~\cite{hackernews-financial-backup} and geopolitical backing~\cite{gaoHinCTICyberThreat2022}. The attackers possess a comprehensive set of malicious skills. Attackers intentionally pick their targets and gather intelligence persistently and stealthily. Cybersecurity practitioners identify these attacks as \textit{Advanced Persistent Threats} (APT) because these attackers have sufficient levels of expertise, capability, and incentives for launching cyberattacks. In 2022, 78\% of companies encountered downtime, and 68\% encountered data loss due to APT attacks~\cite{purplseSec}.  The two following characteristics make APT attacks a severe threat. First, the \textit{mean-time-to-detect} of ongoing APT attacks is reasonably long. IBM found in 2022 that a data breach event goes unnoticed for 197 days on average~\cite{ebroker2022attackstatistics}. Second, APT attackers take the \textit{living-off-the-land} approach. They persistently keep studying their potential victims to circumvent the defense of target organizations. Thus, defending against these attacks requires a deeper understanding of the behavior of the malware used by the adversaries, such as what behaviors are most likely to appear and how the behaviors are connected.

From a technical standpoint, APT attacks are complicated and often combine multiple \textit{adversarial techniques}, a term used to refer to the adversarial behavior or the specific methods used by malware in a cyberattack~\cite{mitreBestPratices, maymi2017towards}. Sending phishing emails containing malicious attachments and exploiting vulnerabilities are two examples of adversarial techniques. Adversaries perform a series of techniques through developing custom malware or using openly available malware. Overall, adversarial techniques primarily reflect the behavior of the adversaries and the malware used by the adversaries. From the defender's point of view, understanding how APT attacks consist of  adversarial techniques - can aid in proactively defending from these attacks. Cybersecurity vendors and analysts publish articles and reports covering step-by-step details of adversarial techniques used by malware in APT cyberattack incidents. These reports and articles are often called \textit{CTI reports} because the reports serve as a source for cyber-threat intelligence (CTI) - actionable information on adversarial techniques that aid in defending organizations from existing or potential threats~\cite{dalziel_introduction_2014, mcmillan_definition_nodate}. 




We are interested in understanding the following from CTI reports: \textit{how adversarial techniques are used to construct APT attacks}. We analyze two specific characteristics of APT attacks: prevalent adversarial techniques; and recurring pairs of adversarial techniques. Prevalent adversarial techniques inform defenders about widespread adversary behaviors. Recurring pairs of techniques inform defenders about how adversary behaviors are connected in the context of cyberattacks. Overall, the analysis captures the structure of APT attacks in terms of adversarial techniques. \textit{The goal of this research is to inform cybersecurity practitioners about how adversaries form cyberattacks through an analysis of adversarial techniques documented in cyberthreat intelligence reports.} We investigate the following research questions (RQs): \textbf{RQ1:} From the empirical analysis of CTI reports, what adversarial techniques are prevalent in APT attacks, and how are they implemented? \textbf{RQ2:} From mining the techniques in CTI reports, what recurring pairs of adversarial techniques appear in APT attacks, and how are they combined?


We propose a systematic methodology to answer the RQs. We use the MITRE ATT\&CK~\cite{attackhomepage} framework, which provides a taxonomy of adversarial techniques and a repository of CTI reports of APT attacks. We systematically represent each attack as a collection of adversarial techniques by using the mapping provided in MITRE ATT\&CK. To answer RQ1, we identify prevalent techniques by analyzing the frequency and trend of the techniques. We then use qualitative coding to characterize the implementation of the techniques in practice (e.g., what actions adversaries perform, what information they gather, and what system component they abuse or affect). To answer RQ2, we identify statistically-significant pairs of adversarial techniques using association rule mining and then perform qualitative coding to characterize the relationships among the techniques in pairs. Overall, we contribute the following: (a) a systematic methodology to identify prevalent adversary behaviors and their association with one another observed in APT attacks, (b) an empirical analysis of the frequency and trend of the adversarial techniques in CTI reports, (c) a qualitative coding process to characterize the implementation of adversarial techniques, (d) an association rule mining-based analysis methodology for identifying recurring pairs of adversarial techniques found in the CTI reports, (e) A qualitative coding process to characterize the relationship types among the techniques in a recurring pair, and (f) a curated dataset of 667 CTI reports from 2008 to 2022 and replicable analysis scripts for independent researchers and practitioners\footnote{see Section 1(H) of the supplementary file}.


We organize the rest of the paper as follows. Section~\ref{sec:concept}, and~\ref{sec:relatedWork} discusses several key concepts and related work. Section~\ref{sec:methodology},~\ref{sec:findingsRQ1}, and~\ref{sec:findingsRQ1} report the methodology and findings. Section~\ref{sec:evaluation} evaluate the findings. Section~\ref{sec:discussion}, and~\ref{sec:threats} highlights several discussion points and threats to the validity. We finally conclude the article in Section~\ref{sec:conclusion}.

\section{Key Concepts}
\label{sec:concept}
We discuss several key concepts related to the study in the following subsections. 


\subsection{MITRE ATT\&CK}

To model real-world cyberattacks through a set of terminologies, the MITRE corporation introduced ATT\&CK in 2013. ATT\&CK models multi-stage cyberattacks through a taxonomy of adversarial techniques used by malware sighted in APT attacks. ATT\&CK catalogs the cybercrime groups and malware along with their set of adversarial techniques. For each case of a group or malware using technique(s), ATT\&CK cites the CTI reports where an instance of a technique usage is documented - which we refer to as a \textit{citation}. The catalog of adversarial techniques gets updated each year, and version 12 is current at the time of writing. To model adversary behavior, ATT\&CK uses a set of terms: \textit{tactics, techniques, and procedures (TTPs)}~\cite{techniqueDefNIST,techniqueDefFeroot}. \textit{Tactic} refers to the adversary’s tactical goal for performing a malicious action. For example, \textit{credential access}~\cite{TA0006} is a tactic for gaining or obtaining credentials of target victim organizations. \textit{Techniques} refer to how an adversary achieves a tactical goal by performing a method~\cite{mitreAttack, attack-design}. An example technique to enable attackers to obtain the \textit{credential-access} tactic is \textit{input capture}. In this technique, attackers collect inputs from victim users and identify inputs that could relate to credentials. \textit{Procedures} refer to the specific implementation of an adversarial technique in malware~\cite{attack-design}.  For example, adversaries can deploy \textit{FlawedAmmyy}~\cite{flawwedAmmy}, for capturing mouse and keyboard events and identifying keyboard inputs regarding credentials. The ATT\&CK repository contains an enumeration of tactics. Each tactic has an enumeration of techniques. Each technique has an enumeration of corresponding procedure(s). ATT\&CK constructs the enumeration of procedures for each technique using the following steps: (a) identify descriptions of how the technique is implemented in an attack from CTI reports; (b) rephrase the description and record the rephrased description as a procedure; (c) cite the URL of the corresponding CTI reports; and (d) attribute the cybercrime groups or malware who performed the procedure. The identifier of a tactic follows the convention \textit{TAxxxx}. The identifier of a technique follows the convention \textit{Txxxx}. A technique may have sub-techniques. Sub-techniques were first introduced in ATT\&CK Version 7. ATT\&CK defines sub-techniques as individual techniques. However, they are called sub-techniques because sub-techniques of a given technique define specific methods based on a relatively abstract method defined in the technique. The identifier of the sub-technique follows the convention of the identifier of a technique, followed by \textit{.xxx}.

\begin{figure}
    \centering
    \includegraphics[width=0.5\columnwidth]{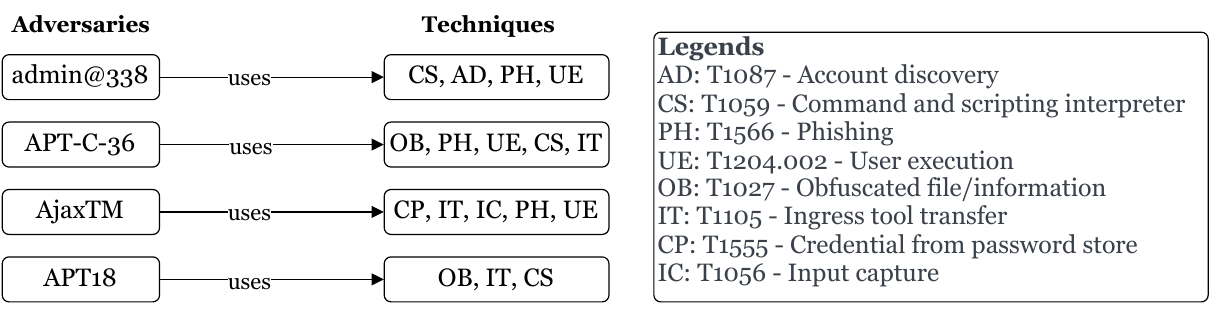}
    \caption{Example of four real-world cyberattacks consisting of multiple techniques~\cite{admin338, aptC36, ajaxtm, apt18}}
    \label{fig:attack-example}
\end{figure}

\subsection{Association Rule Mining}
\label{sec:concepts:amr}
Association Rule Mining (ARM) is a methodology for discovering potential relations among the items in a dataset. Fig.~\ref{fig:attack-example} shows an example of a dataset containing four sets of techniques observed in four real cyberattacks. We refer to each set of techniques as \textit{itemsets}. Fig.~\ref{fig:attack-example} contains four itemsets, where both \textit{PH}, and \textit{UE}  occurred with each other three times. The observation indicates potential relation (e.g., correlation or causation) between these two techniques. The relation between these two techniques can be indicated using the following notation: $PH \implies UE$. The notation denotes that if $PH$ occurs, then $UE$ may also occur. Moreover, this notation mathematically denotes an association rule between these two techniques. An \textit{association rule} looks identical to an \textit{if-then} expression. For example, the example rule: $PH \implies UE$ denotes that \textit{if} PH occurs, \textit{then} UE also occurs. We refer to the \textit{if} portion of the rule as an \textit{antecedent} (e.g., PH), and the \textit{then} portion of the rule as a \textit{consequent} (e.g., UE). ARM takes a parameter named \textit{Minimum Support(minSup)} to extract association rules. \textit{minSup} denotes that the association rule materializes in at least \textit{minSup}\% of data points. We provide examples from Fig.~\ref{fig:attack-example} where we assume ARM is run with $minSup = 0.5$. We identify an association rule: $CS \implies OB$ having $support = 0.5$. \textit{Support} refers to the percentage of data points where the rule holds. As we see: (a) both CS and OB appear in two out of four itemsets, and (b) CS appears in three itemsets and OB in two out of these three itemsets. 

\subsection{Interestingness Measure}
\label{sec:concepts:interestingness}
In an extensive dataset, thousands of association rules can be discovered using ARM. However -  how significant are these discovered rules - needs to be identified. Researchers proposed several interestingness measures that quantify the significance. Piatetsky et al.~\cite{piatetsky1991discovery} proposed the following three desired properties of a good interestingness measure ($M$) of an association rule among two items, $A$ and $B$, denoted by $A \implies B$: \textbf{Property 1:} $M=0$, if $A$ and $B$ are statistically independent, \textbf{Property 2:} $M$ monotonically increases with $P(A,B)$ given that $P(A)$ and $P(B)$ remain unchanged, and \textbf{Property 3:} $M$ monotonically decreases with: (a) $P(A)$ given that $P(B)$, and $P(A,B)$ remain unchanged; and (b) $P(B)$ given that $P(A)$, and $P(A,B)$ remain unchanged. In this study, we use the Phi Correlation Co-efficient~\cite{matthews1975comparison} as an interestingness measure for the discovered association rules among the techniques. The measure satisfies the above-stated three properties, as shown in the study of Tan et al.~\cite{tan2002selecting}.

\section{Related Work}
\label{sec:relatedWork}
This section discusses several studies related to our work. 

\textbf{Extracting information on adversarial techniques from unstructured sources} CTI reports describe past attacks by documenting the adversarial techniques. The reports are written in plain English and are unstructured. Researchers applied natural language processing and machine learning techniques to extract information from unstructured sources. Husari et al. defined an ontology of adversarial techniques and extract techniques from malware reports using unsupervised text similarity~\cite{husari_ttpdrill_2017}. Wu et al. classified sentences into e-commerce-related adversarial techniques using keyword extraction and named entity recognition. Researchers also construct knowledge graphs from the CTI reports. Li et al. and Piplai et al. proposed approaches for constructing knowledge graphs of adversarial techniques and malware indicators from CTI reports by graph alignment and relation classification, respectively~\cite{Li2022AttacKGCT, piplai_creating_2020}. Satvat et al. and Gao et al. constructed provenance graphs of malware behavior from CTI reports. They utilized the graph for two following downstream tasks: malware classification and threat hunting, respectively~\cite{satvat_extractor_2021, Gao2021EnablingEC}. The authors in~\cite{hortanetoCyberThreatHunting2020, gyllingMappingCyberThreat2021, netoPolymerAdaptiveKill2021} utilized knowledge graphs of adversarial techniques and malware indicators for threat-hunting purposes. Rahman et al. systematically surveyed 84 studies on extracting cyberattack-related information from CTI reports using natural language processing and machine learning techniques~\cite{Rahman2022What}. The author pointed out that the existing studies often suffer from low prediction performance. The authors emphasized that, along with automation, a deeper understanding of prevalent threats and their correlation would lead to a better actionability of extracted CTI. Our study focuses on the prevalence and pairings of techniques to enable a comprehensive understanding of the nature of APT attacks.

\textbf{Co-occurrence analysis and probabilistic models of adversarial techniques} In~\cite{noorAssociationRuleMiningBased2018, abuFormulationAssociationRule2021}, the authors performed a co-occurrence analysis of observed adversarial techniques, malware, indicators, and cybercrime groups in cyberattacks. They identified 155 and 81 co-occurring pairs, respectively. These pairs reflect relationships among techniques, indicators, groups, and malware. In~\cite{kimCooccurrenceBasedSecurity2020}, the authors performed a co-occurrence analysis of cyber-physical systems-related attack indicators and visualized their findings. In~\cite{kaiserAttackForecastPrediction2021, al-shaerLearningAssociationsMITRE2020}, the authors performed clustering and dimension reduction to identify similar groups of adversarial techniques observed in cyberattacks. In~\cite{sentunaNovelEnhancedNaive2021}, the authors proposed a probabilistic model of co-occurring techniques, and they used the model to predict future sets of adversarial techniques given a set of observed adversarial techniques. In~\cite{elitzurAttackHypothesisGeneration2019}, the authors modeled the relation between adversarial techniques as a link prediction problem, and the model can predict future sets of adversarial techniques given a set of observed adversarial techniques. In~\cite{samtaniProactivelyIdentifyingEmerging2020}, the authors clustered the cyberattack-related keywords in hacker forums to identify the semantic shift of attack strategies among attackers. The studies above investigated interrelations among adversarial techniques quantitatively through rule mining, probabilistic analysis, graph classification, or clustering. We further investigate interrelation among adversarial techniques both quantitatively and qualitatively and, thus, provide a comprehensive understanding of the characteristics of interrelation among techniques.  


\textbf{Survey and systematic analysis of past cyberattacks} Information related to previous cyberattacks, cyber-criminal groups, and malware is available in the following sources: CTI reports, advisories, malware and vulnerability databases, etc. Collectively, these sources are rich with information on various aspects of cyberattacks. Hence, researchers have used this information to survey and systematically analyze past cyberattacks from diversified perspectives. Researchers investigated whether the lack of security controls is responsible for cyberattacks. For example, Neto et al., and Khan et al., investigated how a data-breach incident\footnote{https://www.capitalone.com/digital/facts2019/} occurred at Capital One bank in 2019 and what lack of security measures were responsible for the incident~\cite{neto2021case, khan2022systematic}. The work demonstrated how systematic analysis of past incidents could improve cybersecurity and provided an investigation framework for individuals and organizations. Researchers aggregated and systematized existing knowledge of past cyberattacks. Lemay et al. surveyed cyber-criminal groups reported in CTI reports and compiled a quick reference of malicious activities of the groups~\cite{lemay2018survey}. In~\cite{shin2022focusing}, the authors investigated social engineering attack surfaces and compiled a set of countermeasures from previous campaigns. In~\cite{edwards2017panning, urban2020plenty}, the authors showed that online and social media employee information provides sufficient details for attackers to launch social engineering campaigns. Villalon et al. compiled a taxonomy of malware delivery and malware persistence-related techniques in~\cite{villalon2022taxonomyDelivery, villalon2022taxonomyPersistence}. Saleem et al. and Oosthoek et al. compiled taxonomies for data-breach attacks and Windows malware, respectively, in~\cite{saleem2020sok, oosthoek2019sok}. Kupyers et al. analyzed 60K cybersecurity incidents across six years of an anonymous organization. They discovered that scrutinizing cyberattack incidents may reduce the impact~\cite{kuypers2016empirical}. In~\cite{alkhalifah2019empirical}, the authors analyzed 65 block-chain attacks from 2011-19. they identified three types of vulnerabilities related to these attacks: byte-code, smart contract, and block-chain network. In~\cite{blakely2022exploring}, the authors analyzed nine breach reports based on four enterprise risk management criteria. They investigated whether the reports could investigate potential limitations of internal controls. They found that reports contain insufficient information for improving the risk profiles of a given entity. They suggested improving the writing and content quality of breach reports and improving the timeliness of reports. Researchers also performed empirical studies on various types of malware. For example, Ugarte-Pedrero et al. analyzed a daily snapshot of malware from a malware database to show how to identify the most representative malware samples from a massive malware population. They proposed a framework for filtering and grouping malware by their behavior to support manual verification~\cite{ugarte2019close}. Botacin et al. performed a longitudinal analysis on 41K Brazilian financial malware from 2012-20~\cite{botacin2021one}. They found that 83\% of the malware was distributed through social engineering platforms. They found adversaries significantly modifying their behavior, such as their file formats and persistence techniques. They advised cybersecurity practitioners to consider the socio-cultural aspects of a given country regarding potential social engineering attack vectors and design countermeasures to address those. Alrawi et al. performed an empirical analysis of 166K internet-of-things (IoT) malware~\cite{alrawi2021circle}. They discovered that IoT malware is different from traditional malware in spreading laterally and communicating with remote botnets. They also emphasized that the large-scale IoT infrastructure is not well-positioned to withstand attacks of sophisticated IoT malware. 

Overall, the studies above analyzed various perspectives of malware seen in past cyberattacks and conveyed actionable takeaways for practitioners. We draw inspiration from the studies above. To our knowledge, no comprehensive research systematically investigates the characteristics of prevalent adversarial techniques and how adversaries pair them. In this study, we address this gap. 

\section{Methodology}
\label{sec:methodology}
In this section, we discuss the methodology to conduct the study. 

\subsection{Construct dataset}
\label{method:dataset}
This section discusses how we construct the dataset for the study. 

\subsubsection{Obtain the initial dataset from MITRE ATT\&CK}
To answer the RQs, we choose the source(s) for obtaining dataset(s).  We select MITRE ATT\&CK as the source for obtaining the dataset for the following reasons. First, MITRE ATT\&CK contains a taxonomy of adversarial techniques and a repository of CTI reports. Second, MITRE ATT\&CK is also well known to cybersecurity practitioners, and MITRE actively collaborates with practitioners to update its taxonomy and repository~\cite{strom2018mitre}. Third, MITRE ATT\&CK repository contains the information on which techniques are reported in which CTI reports. 

\subsubsection{Construct the initial dataset}
We download the STIX\footnote{https://oasis-open.github.io/cti-documentation/stix/intro.html} export file of the MITRE ATT\&CK from their official repository~\cite{attackgithub}. We use Version 12 of the MITRE ATT\&CK repository. The file contains the following: (a) taxonomy information of techniques; (b) CTI reports; and (c) which techniques are reported in which CTI reports. We construct our initial dataset according to the following: (a) a technique aids adversaries in achieving a tactic; (b) techniques include citations; (c) citation provides the URLs of the CTI reports from where ATT\&CK records procedures for the techniques; and (d) CTI report attributes which cybercrime groups or malware are responsible for the described cyberattack. The dataset contains 14 tactics and 594 techniques, including 401 sub-techniques, 10,514 procedures, and 1,070 citations. 
A summary of the initial dataset's tactics, techniques, and procedures is reported in Table~\ref{tab:tacticdescription}.


\begin{table*}[]
    \centering
    \scriptsize
    \caption{The fourteen tactics cataloged in MITRE ATT\&CK}
    \label{tab:tacticdescription}
    \begin{tabular}{lp{7cm}rr}
        \toprule
        \textbf{Tactic} & \textbf{Description} & \textbf{\# Techniques} & \textbf{\# Procedures} \\ \midrule
        
        TA0043: Reconnaissance & Adversary plans to conduct future operations & 43 & 80 \\
        TA0042: Resource development & Adversary obtains resources for conducting cyberattacks & 43 & 219 \\
        TA0001: Initial access & Adversary gains entry to a victim system & 12 & 263 \\
        TA0002: Execution & Adversary executes malicious instructions in a victim system & 33 & 1,389 \\
        TA0003: Persistence & Adversary remains hidden in a victim system & 69 & 663 \\
        TA0004: Privilege escalation & Adversary performs malicious actions with higher privilege & 24 & 128 \\
        TA0005: Defense evasion & Adversary bypasses victim system's defense & 150 & 2,392 \\
        TA0006: Credential access & Adversary obtains credentials of victim systems & 56 & 373 \\
        TA0007: Discovery & Adversary obtains information of a victim system & 37 & 2,010 \\
        TA0008: Lateral movement & Adversary propagates to connected systems to a victim system & 15 & 187 \\
        TA0009: Collection & Adversary collects data and information from a victim system & 33 & 870 \\
        TA0011: Command and control & Adversary establish communication among a victim system and adversary-controlled remote systems & 36 & 1,518 \\
        TA0010: Exfiltration & Adversary steals away data and information from a victim system & 17 & 239 \\
        TA0040: Impact & Adversary disables, disrupts, deter or destroy operations of a victim system & 26 & 183 \\ \midrule

        \multicolumn{4}{c}{Definitions and examples of all tactics, and techniques can be found in~\cite{alltactics}, and~\cite{alltechniques} respectively} \\
        
    \bottomrule
    \end{tabular}
\end{table*}


\subsubsection{Construct a mapping between techniques and CTI reports}
In the initial dataset, the techniques have citations, and the citations cite CTI reports. We construct a mapping between the CTI reports and the techniques. As a result, we get information on the set of techniques described in the report. For each report in the initial dataset, we record the mapping information. Based on the mapping, we assume that each cyberattack described in the corresponding CTI report is represented by a set of techniques mapped to the CTI report.



\subsubsection{Filter citations from the initial dataset}
Not all the citations in the initial dataset describe cyberattack incidents. Hence, we establish inclusion and exclusion criteria for filtering the citations so that we only have citations for CTI reports that describe cyberattack incidents. The inclusion criteria are: (a) the CTI report must be in English; (b) the CTI report must be accessible online; (c) the CTI report describes actual cyberattack incidents step by step, and (d) the CTI report maps at least two techniques. The exclusion criteria are: (a) the cited URL is detected by the browser as insecure; (b) the content of the URL does not contain any cyberattack description; and (c) the cited URL is a social media post, definition of a cybersecurity concept (e.g., Wikipedia), API documentations, source code repositories, etc. After applying these two sets of criteria, we obtain 920 citations. 


\subsubsection{Identifying citations reporting the same cyberattack incident}
Multiple cybersecurity vendors may report cyberattacks on reputed organizations, especially in the case of high-profile cyberattacks. Hence, the initial dataset may have multiple citations of CTI reports covering the same cyberattack incident. We call these CTI reports as duplicate CTI reports and the corresponding citations as duplicate citations. We need to remove the duplicate CTI reports to quantify the prevalence of techniques. We first identify the set of all pairs of duplicate citations. We then combine the pairs of duplicate citations as a single unique citation. We assume each unique citation covers a unique cyberattack incident.  We use the following criteria to identify if two reports are \textit{duplicate}. \textbf{Criterion-1: The two reports must attribute at least one common cybercrime group or malware.} For example, two CTI reports from~\cite{wellmainCISA} and~\cite{wellmailNCSC} describe the same malware named \textit{WellMail}\footnote{https://attack.mitre.org/software/S0515/}. We manually checked these reports and found that these two CTI reports discuss the same cyberattack. However, this criterion may incorrectly identify two reports describing different incidents. For example, in the initial dataset, we identify a cybercrime group named \textit{FIN7}\footnote{https://attack.mitre.org/groups/G0046/}, which has been attributed in a CTI report from 2017~\cite{fin72017restaurant} and a CTI report from 2022~\cite{fin72022USRansomware} for two separate cyberattacks. \textbf{Criterion-2: If the two reports are the same, they should have a smaller difference between their publication dates than that of two other reports discussing different cyberattacks.} Consider the two reports discussing \textit{FIN7} as pair-A, and the two reports discussing \textit{WellMail} as pair-B. The difference between the publication date of pair-A is five years. However, the difference between the publication date of pair-B is one month. Five years of date difference is a relatively long period. However, one month of difference is a relatively more reasonable difference in case of two reports covering the same incident. Thus, we assume the two reports are \textit{duplicate} if both criteria hold. However, we need to establish a threshold ($\tau$), difference for criterion-2. We perform the following steps.

\textbf{Step-1: Identify publication date:} We identify the publication date mentioned in each citation. MITRE ATT\&CK documents the publication dates for most of the citations. However, we also identify citations with no publication date mentioned in the ATT\&CK dataset. We manually open the URL in the browser of each citation and collect the publication date mentioned in the corresponding CTI report. If the report does not mention any date, we use the following to infer the publication date: (a) we first check the URL to see if the URL pattern indicates any date, such as \verb|www.example.com/2022/10/31/report.html|; (b) if (a) does not work, we then search the URL in Google search engine\footnote{https://www.google.com/} to identify its upload or first appearance date; (c) finally, if (a) and (b) does not work, we remove the citation from the initial dataset. We removed 8 citations from the initial dataset.

\textbf{Step-2: Find pairs of reports having a common group or malware:} Step-2 implements criterion-1. We identify pairs of citations such that the two reports from the two citations in a pair attribute at least one common cybercrime group or malware. The ATT\&CK repository contains the information of each citation attributing cybercrime groups or malware responsible for the cyberattack. A citation can attribute multiple groups or malware. Hence, we construct a pair of citations if the following condition holds: at least one group or malware exists in the intersection of the sets of attributed groups or malware from the two corresponding CTI reports. We identify 2,425 such pairs in the initial dataset. We call these pairs $P = \{p_1, p_2, p_3, ..., p_{2425}\}$. Each pair is defined as $p_i = (c_1, c_2)$, where $c_1$ and $c_2$ cite two reports attributed to at least one common cybercrime group or malware.  

\textbf{Step-3: Determining the value of $\tau$:} 
Step-3 implements criterion-2. The step determines the value of the threshold $\tau$ empirically. We first take random samples of pairs having different publication date differences (i.e., pairs differing by one month, pairs differing by two months, and so on). We then manually check the fraction of duplicate pairs for each date difference. We then plot the fraction against the date differences and apply the elbow method~\cite{thorndike1953belongs} to determine $\tau$ visually. We report the steps: \textbf{Step 3a:} We set \textit{month} as a unit for the time difference of publication dates. We assume a month equals to 30 days in the context of this study. \textbf{Step 3b:} We select $n$ subsets of $P$, referred to as: $SP_i \subset P $, where $i=1,2,3,...,n$. $SP_i$ holds the following condition. The publication date difference ($\delta$) among any two citations in a member of $SP_i$ is $i-1 < \delta \leq i$. For example, the publication date difference between the two members of $SP_2$ is more than one month, but at most, two months. The publication date difference between any two members of $SP_3$ is more than two months, but at most three months. \textbf{Step 3c:} We randomly construct a subset ($RSP_i$) from each of $SP_i$. For example, $RSP_1 \subset SP_1$ and $RSP_2 \subset SP_2$. Each of $RSP_i$ has $s$ number of members. \textbf{Step 3d:} We manually check the corresponding CTI reports from each pair of $RSP_i$. We then count the percentage of pairs ($r_i$), describing the same incident among out-of $s$ pairs in $RSP_i$. \textbf{Step 3e:} We apply the elbow method to identify the value of $i$ where the first big drop in the $r_i$ is observed. We then set the threshold to $\tau$ to $i$ unit of time. \textbf{Step 3f:} In this study, we set $n=5, s=20$. After applying the elbow method, we find $r_1 = 0.95, r_2= 0.85,$ and $ r_3 = 0.15$. Thus, we identify $\tau = 2$.


\textbf{Step-4: Merging the citations:} We detect all pairs satisfying criterion-1. and criterion-2. We then construct a graph of citations where two citations have an edge between them if they are members of the same pair. Then, we compute the graph's number of \textit{connected components}. We assume all the citations in a single connected component discuss the same cyberattack incident. We merge the citations of the same connected component as a single citation. Hence, the total number of unique cyberattacks is the count of total connected components. We then create a set of techniques by computing the union of the list of techniques mapped to each corresponding report from the citation of the same connected component.


\subsubsection{Dataset Summary}
\label{sec:method_dataset_summary}
After performing the four steps, we obtained 667 citations, including 146 merged citations from 2008-2022\footnote{The list of 667 CTI reports are mentioned in Section 1(C) of the supplementary file}. Each of the citations cites a CTI report, and thus, we have a set of 667 CTI reports. We randomly selected 25\% (161) of the CTI reports to gain insight on the metadata information of the CTI reports\footnote{The list of 161 reports are reported in Section 1(B) of the supplementary file}. We read the title and introduction and took notes of the following properties: source, report category, malware types, attack types, and impacted sector. We report the top five values for each property in Table~\ref{tab:reportAttributes}. The table indicates that cyberattacks described in our dataset are primarily launched against the government, defense, and critical sectors like energy, media, education, etc. APT actors launched all of the cyberattacks. The motive of the attacks is mainly stealing data and credentials for espionage motives. The reports primarily analyze cybercrime groups and malware activities. Top reported malware aids adversaries in gaining remote access, exfiltrating data through backdoors, and asking for ransoms. Publishers of these CTI reports are reputed cybersecurity vendors. These sampled reports demonstrate the nature of the APT attacks: targeted for geopolitical and economic reasons, exfiltrate sensitive data, and ask for ransomware. Each report in the dataset documents the use of adversarial techniques. We refer to the documentation of a technique in a CTI report as a mention. In the dataset, the 667 CTI reports document the use of 10,370 adversarial techniques in total - which we refer to as the total mentions of techniques across all CTI reports. The average and median number of techniques documented per report are 15.59 and 13, respectively. We identify 452 out of 594 cataloged techniques in ATT\&CK are documented by at least one report, and rest of the 142 techniques are not mentioned in any CTI report in our dataset.



\begin{table*}
    \centering
    \scriptsize
    \caption{The top five values of properties of randomly sampled reports. The numbers in the parenthesis denote counts}
    \label{tab:reportAttributes}
    \begin{tabular}{ll}
    \toprule
    
    \textbf{Properties} & \textbf{Top five values} \\ \midrule

    Domain & Government (32), defense (10), energy (11), education (8), media (7) \\

    Attack types & Espionage (19), data theft (16), credential theft (7), phishing campaign (6), supply chain attack (2) \\ 

    Report category & Cybercrime group analysis (87), malware analysis (68), alert (2), legal report (1) \\ 

    Malware types & Remote access tool (16), ransomware (14), backdoor (12), trojan (9), spyware (6) \\

    Publisher & Palo Alto Networks (22), ESET (17), Talos Intelligence (11), Kaspersky (9), Symantec (7) \\
    
    \bottomrule
    \end{tabular}
\end{table*}


\subsection{Methodology for RQ1}
\label{rq1Methodology}
To answer\textit{ RQ1: From the empirical analysis of CTI reports, what adversarial techniques are prevalent in APT attacks, and how are they implemented?}, we first identify the prevalent adversarial techniques. Then, we characterize the implementation of these techniques. We first identify prevalence by analyzing two pieces of information computed from the dataset related to techniques: frequency and trend. 
We then perform qualitative coding to characterize the implementation of prevalent techniques. Below, we report the steps toward RQ1.


\textbf{Step-1: Analyze the frequency of techniques:} We first count the frequency of all techniques, including sub-techniques, across all CTI reports. We treat sub-techniques as individual techniques. The frequency of a technique refers to the count of reports a technique is mapped to. For example, if a technique $te$ is mapped to $n$ reports in the dataset, the frequency of $te$ is $n$. The frequency of a technique indicates to what extent the technique is prone to be used by adversaries. 

\textbf{Step-2: Analyze the trend of techniques:} We perform a Mann Kendall's test~\cite{mann1945nonparametric} to identify potential increasing or decreasing trends in the usage of techniques in the past five years (i.e., from 2018-2022, at the time of writing). We perform the test for each technique in the dataset. First, we identify the number of reports mentioning each year's techniques in the chosen range of years for each technique. Then we divide the count by the total number of reports published each year in the range. We then use the \verb|pymannkendall| \textit{python} package\footnote{https://pypi.org/project/pymannkendall/} to perform the trend test. For each technique, the test indicates whether a statistically significant trend is observed for a given technique in the given year range. The test outputs one of the following three trend information for each technique: \textit{increasing trend}, \textit{decreasing trend}, and \textit{no trend}. 

\textbf{Step-3: Construct a frequency-trend matrix:} We sort the techniques by their respective frequency in descending order. We then divide the techniques into three groups according to their frequency: \textit{high-frequency} (top 33 percentiles), \textit{medium-frequency} (34 to 67 percentiles), and \textit{low-frequency} (67 to 100 percentiles). Then, we divide all the techniques into three groups according to their trend: \textit{increasing}, \textit{no trend}, and \textit{decreasing}. We create a three-by-three matrix, where each technique will be exclusively placed into one of nine cells according to their group in frequency and trend. Table~\ref{tab:matrix} shows the matrix. For example, the top-left cell of the table would contain the list of techniques that are of both high frequency and increasing trend.



\textbf{Step-4: Identify the prevalent techniques using the matrix:} The matrix shown in Table~\ref{tab:matrix} represents a technique in two dimensions: frequency and trend. We characterize the prevalence of techniques based on the technique's position in the matrix. The matrix demonstrates how much a technique is frequently used and trendy among adversaries. The techniques placed in the \textit{top-right} corner are highly frequent and on an increasing trend. Observing the properties of the four corners, we assume that the \textit{top-right} corner represents prevalence most because techniques on the cell are highly frequent, and their frequency is expected to increase in the future. Hence, we call a technique \textbf{prevalent} if the technique is in one of the three following cell categories, shown in italics in Table~\ref{tab:matrix}: (a) high/increasing, (b) high/no trend, and (c) medium/increasing. Note that we assumed categories (b) and (c) as prevalent in addition to (a) for the following reasons. Techniques in (b) are highly frequent and are on \textit{no-trend} and, thus, are expected to remain highly frequent. Techniques in (c) are of medium frequency, but they are on an increasing trend and, thus, expected to be highly frequent in the future. Moreover, we do not consider techniques in the \textit{bottom-right} corner, even though those techniques are highly frequent. The reason is those techniques are on decreasing trend and are not expected to be highly frequent in the future. 

\textbf{Step-5: Characterize the implementation of prevalent techniques:} Adversarial techniques are often abstract and described theoretically. However, procedures provide information on the specific implementation of the techniques. We first obtain the set of procedures from the CTI reports in the dataset mapped to the prevalent techniques. The first two authors apply qualitative coding to characterize the implementation of techniques by the following steps. \textbf{Step-5.1: Apply labels} The authors read the procedures and apply one or more labels to them. The labels are primarily verbs and nouns in the procedure sentences. The labels reflect specific context and additional information in the procedure description. \textbf{Step-5.2: Open coding} The authors then apply codes to the labels using the open coding technique~\cite{saldana_coding_2015}. \textbf{Step-5.3: Axial coding} The authors categorize the codes obtained in Step-5.2, using axial coding~\cite{saldana_coding_2015}. \textbf{Step-5.4: Inter-rater agreement} The first and the second author perform an inter-rater agreement on axial codes and resolve their disagreements.

\begin{figure}
    \centering
    \includegraphics[width=\textwidth]{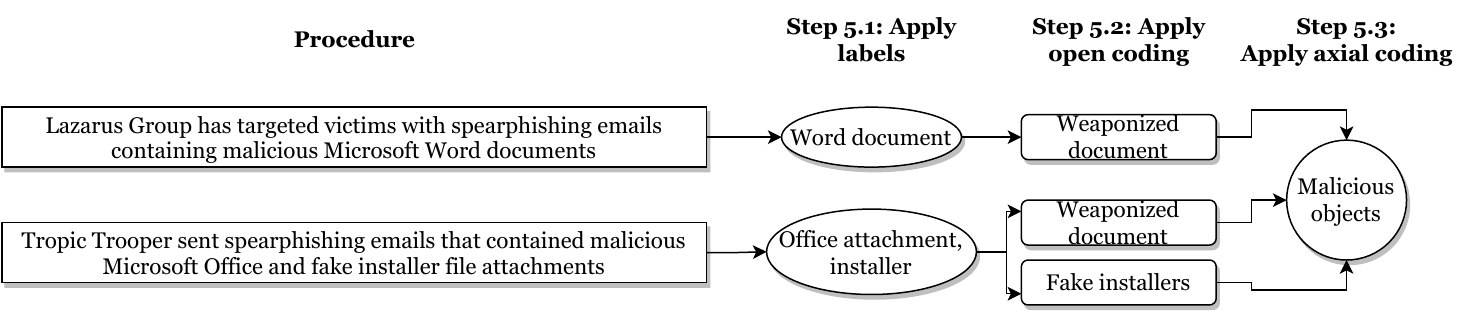}
    \caption{Example of the qualitative coding on the procedure}
    \label{fig:procedureCode}
\end{figure}

In Fig.~\ref{fig:procedureCode}, we show an example where we show two procedures for \textit{T1566.001: Spearphishing attachment}. In the first procedure, the attachment was a document file, and we applied a label \textit{word document} to the procedure in Step-5.1. In the second procedure, we applied \textit{office attachment}, and \textit{installer} as two labels to the procedure in Step-5.1. In Step-5.2, we applied the code \verb|Weaponized document| to both labels: \verb|Word document|, and \verb|Office attachment|, as the two labels represent the same concept. Then, in Step-5.3, we group the three identified labels as \textit{malicious objects} category because all three labels are artifacts that can deliver malware to a victim system.


\begin{table*}
    \centering
    \scriptsize
    \caption{The Frequency-Trend Matrix}
    \label{tab:matrix}
    \begin{tabular}{c|c|c|c}
    \toprule
         {} & \multicolumn{3}{c}{\textbf{Frequency}} \\ \midrule
         \multirow{3}{*}{\textbf{Trend}} & Increasing/Low & \textit{Increasing/Medium} & \textit{Increasing/High} \\ \cmidrule{2-4}
         {} & No trend/Low & No trend/Medium & \textit{No trend/High} \\ \cmidrule{2-4}
         {} & Decreasing/Low & Decreasing/Medium & Decreasing/High \\
    \bottomrule    
    \end{tabular}
\end{table*} 

\subsection{Methodology for RQ2}
\label{rq2:methods}
To answer \textit{RQ2: From mining the techniques in CTI reports, what recurring pairs of adversarial techniques appear in APT attacks, and how are they combined?}, the recurring pairs of techniques need to be identified. We perform the following steps below.


\textbf{Step-1: Perform association rule mining:} After performing Step-4 in Section~\ref{method:dataset}, we identify 667 unique citations. Each of these citations cites a CTI report documenting a set of adversarial techniques used in cyberattacks. Thus, our dataset has 667 itemsets of techniques, which we call \textit{technique-sets}. Thus, we have 667 \textit{technique-sets}. We then perform association rule mining on the 667 \textit{technique-sets}. We set the $minSup = 0.005$ while performing the rule mining, as suggested in~\cite{tan2002selecting}. We identify 13,622 rules at this stage. Each rule contains one antecedent technique ($te_a$) and one consequent technique ($te_c$), such as $te_a \implies te_c$ such that given the antecedent technique occurs, the consequent technique is likely to occur as well (see Section~\ref{sec:concepts:amr}). 

\textbf{Step-2: Filter association rules by interestingness measure}: The obtained rules might be spurious and not statistically correlated. Hence, we filter the spurious rules by computing Pearson Phi co-efficient interestingness measure and Chi-square statistical significance test on the obtained rules. We only keep rules with a 0.20 or more Phi Correlation Coefficient score and statistical significance at $\alpha = 0.05$. We obtain 425 rules at this point. We refer to these 425 rules as the \textbf{recurring pairs of adversarial techniques}. 

\textbf{Step-3: Characterize the relationship among techniques in a pair}: The obtained 425 pairs in Step-2 are significantly correlated, suggesting the two techniques in each pair may have a relationship contributing to the correlation. To characterize the relationship, we perform open coding for each pair~\cite{saldana_coding_2015} to capture the underlying reason contributing to the significant correlation and/or statistical dependence. The first and the third author perform open coding to characterize the relation by (a) reading five randomly-selected CTI reports mapped to both techniques in the pair and (b) taking a short note capturing the relationship. Then the two authors perform an inter-rater agreement and resolve their disagreements. 

\textbf{Step-4: Perform network analysis}: We then construct a graph of techniques where two techniques have an edge between them if the techniques appear in the same pair in Step-3. Depending on the nature of the relation, the graph can be undirected or directed. In the case of undirected graphs, we compute the degree centrality ($\delta$) score of the techniques to investigate how techniques are interrelated with one another across identified relationship types. $\delta$ denotes to what extent a node in a graph is connected to other nodes in the same graph. A higher value in $\delta$ denotes that the corresponding node has a relatively higher number of neighbors (i.e., connected nodes). In the context of this study, an adversarial technique with a higher $\delta$ implies that the technique appears in a relatively higher number of pairs. From the adversaries' perspective, a technique with a higher $\delta$ means the technique is contextually related to a greater number of different techniques. We provide the following example. Assume four adversarial techniques: $T_1, T_2, T_3, T_4$, and five pairs: $(T_1, T_2),(T_1, T_3),(T_1, T_4),(T_2, T_3),(T_3, T_4)$. The degree centrality ($\delta$) of $T_1$ is $\frac{3}{4}=0.75$. Similarly, $\delta$ of $T_2$ is $\frac{2}{4}=0.5$. We compute the in-degree centrality ($\delta_i$) and out-degree centrality ($\delta_o$) scores in the case of directed graphs. A technique having a relatively higher value in $\delta_i$ denotes that the technique is connected to a relatively higher number of other techniques by incoming edges. A technique having a relatively higher value in $\delta_o$ denotes that the technique is connected to a relatively higher number of other techniques by outgoing edges. We provide the following example. Assume four adversarial techniques: $T_1, T_2, T_3, T_4$, and five pairs: $(T_1, T_2),(T_1, T_3),(T_1, T_4),(T_2, T_3),(T_3, T_4)$. $T_3$ has two incoming edges. Hence, $\delta_i$ of $T_3$ is $\frac{2}{4} = 0.5$. $T_1$ has three outgoing edges. Hence, $\delta_o$ of $T_1$ is $\frac{3}{4} = 0.75$.

\begin{table*}
    \centering
    \scriptsize
    \caption{Technique count, median frequency, and percentage of total mentions in the frequency-trend Matrix}
    \label{tab:matrixCount}
    \begin{tabular}{lrrr}
    \toprule
         \multirow{2}{*}{\textbf{Trend}} & \multicolumn{3}{c}{\textbf{Frequency}} \\ \cmidrule{2-4}
         {} & \textbf{Low} & \textbf{Medium} & \textbf{High} \\ \midrule
         Increasing & 21 (1.4) (2.4) & \textit{4 (18.3) (4.1)} & \textit{2 (32.8) (4.2)} \\ \midrule
         No Trend & 508 (0.3) (30.4) & 41 (9.9) (28.9) & \textit{13 (31.0) (29.0)} \\ \midrule
         Decreasing & 5 (3.6) (1.0) & 0(0.0)(0.0) & 0(0.0)(0.0) \\
    \bottomrule    
    \end{tabular}
\end{table*} 

\section{Findings from RQ1}
\label{sec:findingsRQ1}
This section discusses the findings for the \textbf{RQ1}. Section~\ref{rq1:ftm}, and~\ref{rq1:prevalentTechniques} discuss the prevalent techniques. Section~\ref{rq1:implement} discusses the implementation of the identified prevalent techniques.

\subsection{The frequency-trend matrix}
\label{rq1:ftm}
Performing Steps 1-4 reported in Section~\ref{rq1Methodology}, we obtain the frequency-trend matrix, which we show in Table~\ref{tab:matrixCount}. The table shows the techniques count in each of the nine cells in the frequency-trend matrix. For example, the table shows two techniques with high frequency and increasing trend cells (top-right corner). The cell shows these two techniques are reported in 32.8\% of the CTI reports at the median. The cell also shows these two techniques are responsible for 4.2\% of total mentions (see Section~\ref{sec:method_dataset_summary}) of techniques across all CTI reports. We report our observations below, where we only discuss the numerical distribution of techniques based on the matrix\footnote{The list of techniques and their position in the prevalence matrix is reported in Section 2 of the supplementary file}. We discuss the specific techniques in Section~\ref{rq1:prevalentTechniques}. 


Overall, from \textit{medium}, and \textit{high} columns, we see that 66.2\% of the mentions come from a small portion (60 out of 594, ~10\%) of all cataloged techniques in MITRE ATT\&CK. We observe 15 techniques in the \textit{high} column, where these techniques account for 33.2\% of all mentions. Moreover, these techniques appear on a median in around 31\% of all CTI reports. None of the techniques from the high or medium columns are on decreasing trend. \textbf{The observation indicates that, although cybercrime groups and malware often change their operations, the attackers implement a small set of techniques to launch cyberattacks. In this study, we find 60 such techniques.} These techniques pose the most likely threat to an organization. Table 1 of the supplementary materials reports the list of all these techniques. From the \textit{trend} rows, we observe that 562 (88.3\%) of techniques do not show any significant change in their trend. We observe 27 techniques in the increasing trend and 5 in the decreasing trend. Among the techniques in increasing trend, 21 techniques of them are infrequent. We manually inspected the 21 techniques and identified seven from the TA0005: Defense Evasion tactic. \textbf{The observation suggests no technique becomes completely obsolete. However, as the observation from the trend suggests, adversaries can change their attack strategy in terms of how they choose to select TA0005: Defense Evasion-related techniques.}  

\textbf{The frequency-trend matrix's four corners provide specific and actionable suggestions for practitioners.} The two techniques on \textit{top-right} are the most likely to be used by adversaries. Organizations should actively monitor the indicators of these techniques. The 21 techniques on \textit{top-left} are not very likely but are expected to be more prevalent. Organizations should investigate these techniques and deploy countermeasures. The five techniques on \textit{bottom-left} are infrequent, and losing trend. Organizations can expect these techniques will rarely be used. Finally, no techniques exist in the \textit{bottom-right}. From a defender's point of view, the observation is significant because no widespread technique is on a decreasing trend. Thus, attackers keep adding new techniques to their skill sets, signaling defenders to adopt more and more countermeasures.
\begin{table}[]
    \centering
    \scriptsize
    \caption{The prevalent techniques}
    \label{tab:prevalentTechniques}
    \begin{tabular}{lllrl}
    \toprule
    \textbf{Technique Id} & \textbf{Technique Name} & \textbf{Tactic} & \textbf{\% Reports} & \textbf{Frequency/Trend} \\
    \midrule
    T1105     & Ingress Tool Transfer                  & TA0011: Command and Control & 51.72 & High/No trend \\
    T1082     & System Information Discovery           & TA0007: Discovery           & 46.48 & High/No trend \\
    T1027     & Obfuscated Files or Information        & Defense-evasion     & 44.83 & High/No trend \\
    T1071.001 & Web Protocols                          & TA0011: Command and Control & 41.98 & High/No trend  \\
    T1059.003 & Windows Command Shell                  & TA0002: Execution           & 40.03 & High/No trend   \\
    T1083     & File and Directory Discovery           & TA0007: Discovery           & 35.53 & High/Increasing \\
    T1547.001 & Registry Run Keys / Startup Folder     & TA0003: Persistence         & 31.93 & High/No trend  \\
    T1070.004 & File Deletion                          & Defense-evasion     & 31.18 & High/No trend \\
    T1057     & Process Discovery                      & TA0007: Discovery           & 31.03 & High/No trend \\
    T1140     & Deobfuscate/Decode Files or Information & Defense-evasion    & 30.28 & High/Increasing \\
    T1016     & System Network Configuration Discovery & TA0007: Discovery           & 28.49 & High/No trend \\
    T1204.002 & User Execution: Malicious File                         & TA0002: Execution           & 28.04 & High/No trend  \\
    T1566.001 & Spearphishing Attachment               & Initial-access      & 26.84 & High/No trend \\
    T1059.001 & PowerShell                             & TA0002: Execution           & 24.89 & High/No trend \\
    T1033     & System Owner/User Discovery            & TA0007: Discovery           & 24.14 & High/No trend \\
    T1005     & Data from Local System            & TA0009: Collection           & 20.69 & Medium/Increasing \\
    T1106     & Native API                   & TA0002: Execution           & 19.64 & Medium/Increasing  \\
    T1041     & Exfiltration Over C2 Channel & TA0010: Exfiltration        & 16.94 & Medium/Increasing  \\
    T1071.004 & DNS                          & TA0011: Command and Control &  6.75 & Medium/Increasing \\
      
    \bottomrule
    \end{tabular}   
\end{table}

\subsection{The prevalent techniques}
\label{rq1:prevalentTechniques}
Table~\ref{tab:prevalentTechniques} shows the list of 19 prevalent techniques identified from the frequency-trend matrix, as shown in Table~\ref{tab:matrix}. Each of the top ten techniques is reported in at least 30\% of all CTI reports. The prevalent techniques, as shown in italics in Table~\ref{tab:matrixCount} constitutes 37.3\% of total mentions. 


\textbf{The highest number of prevalent techniques comes from the \textit{TA0007: Discovery} tactic}. From the table, we identify five techniques from the tactic. Through the five techniques, adversaries gather information on the victim's: (a) operating system and hardware (\textit{T1082}), (b) local files (\textit{T1083}), (c) running processes (\textit{T1057}), (d) network configuration (\textit{T1016}), and (e) user accounts, groups, and permissions (\textit{T1033}). Adversaries benefit from these techniques by identifying the nature of a target system and what malware and tools they would operate on to compromise the victim. \textbf{The second highest number of prevalent techniques comes from the \textit{TA0002: Execution} tactic.} From the table, we identify four techniques from the tactic. Two of these four techniques are the execution of arbitrary commands through Windows command shell (\textit{T1059.003}) and Powershell (\textit{T1059.001}). However, we also observe adversaries can lure users into executing malicious code on adversaries' behalf (\textit{T1204.002}). In malware, adversaries also abuse native APIs (\textit{T1106}) to perform malicious tasks. 

\textbf{The third highest number of prevalent techniques comes from both the \textit{TA0011: Command and Control}, and the \textit{TA0005: Defense Evasion} tactics}. We identify three techniques from both of the tactics. The table shows that the most prevalent technique comes from the \textit{TA0011: Command and Control} tactic, which is transferring malware from the adversaries' end to the victim's end (\textit{T1105}). Adversaries also establish a command-and-control (C2) channel to send malicious commands and retrieve the data back to the adversaries' end from the victim's end. Adversaries use two application layer protocols: web protocols (\textit{T1071.001}) and DNS (\textit{T1071.004}) to transmit commands and exfiltrate collected data. The table shows that the most prevalent technique from the \textit{TA0005: Defense Evasion} tactic is obfuscating malicious files and information (\textit{T1027}). At the end of their campaign, adversaries delete all their traces (\textit{T1070.004}) to deter forensic analysis. The table also shows the deobfuscation technique (\textit{T1140}) is also prevalent, which aids adversaries in restoring the obfuscated files to their original state. 

\textbf{Only one prevalent technique exists from the \textit{TA0003: Persistence} tactic}. The table shows that adversaries persist their malware by abusing Windows registry keys or Windows Startup folder (\textit{T1547.001}). Through the technique, the malware can automatically run when the system starts. Finally, the table shows two techniques, one from the \textit{TA0009: Collection} and one from the \textit{TA0010: Exfiltration} tactic. Through the two techniques, adversaries collect sensitive data from local systems (\textit{T1105}) and then exfiltrate the data through the C2 channel (\textit{T1041}).  

\textbf{We observe that prevalent techniques reflect the stealthy nature of the APT attacks}. For example, we observe that the top three techniques appear benign if viewed individually, but collectively, the techniques propagate the attack. In \textit{T1082}, adversaries perform OS fingerprinting to gain an understanding of their targeted computing environment. Adversaries then deliver malware (\textit{T1105}) operable in the target environment. The downloaded files and the network traffic are obfuscated (\textit{T1027}), so the content cannot be checked. \textbf{Overall, the detection of the prevalent techniques is challenging. Because they look identical to the benign system operations. Hence, singling out a malicious event from many is difficult.} For example, opening a file is a basic system operation - a user or a legitimate program can open hundreds of files daily. Whether a file is opened for a malicious purpose (e.g., reading a sensitive Word document) is difficult to identify.

\textbf{We observe four types of system feature abuse from the prevalent techniques.} First, email attachments can be abused to lure victim users into initiating a compromise by clicking phishing attachments. Second, communication protocols such as HTTP/S, DNS, FTP, and SSH can be abused for downloading and exfiltrating data. Third, built-in system features, such as command shell, Powershell, and registry run keys, are abused to execute malicious tasks in background. Fourth, we observe five techniques where basic information of system telemetry and system metadata are abused. For example, obtaining the operating system upgrade information is not malicious and does not require a root permission. However, the information can be helpful for an attacker to decide whether the system has an unpatched vulnerability. Security practitioners must identify how adversaries can be prevented from accessing basic system information that can be abused against the organizations.


\textbf{We observe that three prevalent techniques are targeted for Windows OS. We speculate that the attacks are especially targeted at desktop computer users.} \textit{T1059.001: Powershell}, and \textit{T1059.003: Windows command shell} are Windows-specific features. On the other hand, the \textit{T1547.001: Registry run} keys technique uses Windows Registry. Running programs in the background and automatically starting them are also desktop-specific features. Overall, the prevalent techniques suggest desktop environments are the primary target of adversaries.  

\begin{table}[]
\centering
\scriptsize
\caption{Identified open codes in the \textbf{Action} axial code}
\label{tab:axialAction}
\begin{tabular}{p{3cm}rp{5cm}p{6cm}}
\toprule
\textbf{Open codes} &
  \textbf{Count} &
  \textbf{Selected associated labels} & \textbf{Techniques} \\
  \midrule

Apply obfuscation algorithms &
  584 & encode, encrypt, archive, compress, password-protect, code obfuscation, steganography & 
  T1059.001: PowerShell,   T1059.003: Windows Command Shell, T1027: Obfuscated files or information,   T1106: Native API, T1140: De-obfuscation \\ 
  
Deploy malware to the victim system &
  508 & download, retrieve, update, upgrade, copy, transfer &
  T1566.001: Spearphishing Attachment, T1059.001: PowerShell, T1033: System Owner or User Discovery, T1059.003: Windows Command Shell,   T1204.002: Malicious File, T1105: Ingress Tool Transfer, T1071.001: Web   Protocols \\ 
  
Set Registry &
  167 & registry, registry key, registry run key, create registry key, modify registry value & 
  T1059.003: Windows Command Shell, T1059.001: PowerShell,   T1547.001: Registry Run Keys \\ 

Trick user to open documents and enable macro/content &
  152 & enable content, enable macro & 
  T1204.002: Malicious File \\ 
  
Clean malware   footprint &
  126 & delete malware, post compromise cleanup, suicide script, uninstall malware &
  T1059.001: PowerShell, T1083: File and Directory Discovery, T1059.003: Windows Command Shell, T1070.004: File Deletion, T1106: Native   API \\ 
  
  Exfiltrate data &
  95 & exfiltratde to C2, upload to C2, send to C2 & 
  T1041: Exfiltration Over C2 Channel, T1059.001: PowerShell, T1071.001: Web Protocols \\ 
  
Delete local and  system files &
  81 & delete local files, delete system files, overwrite files &
  T1070.004: File Deletion \\ 
  
Modify system and software configuration &
  74 & Modify software configuration, change file permission & 
  T1059.003: Windows Command Shell, T1059.001: PowerShell,   T1547.001: Registry Run Keys \\ 
  
Search and  collect victim system information &
  46 & discovery, reconnaissance, collection, scan & 
  T1566.001: Spearphishing Attachment, T1059.001: PowerShell, T1059.003: Windows Command Shell, T1057: Process Discovery, T1083: File   and Directory Discovery \\ 

Execute in-memory      & 22 &
execute in-memory, fileless execution &
T1059.003: Windows Command Shell, T1059.001: PowerShell,   T1140: De-obfuscation, T1105: Ingress Tool Transfer \\ 
  
Manipulate process &
  11 & create process, kill process, inject process &
  T1057: Process Discovery, T1106: Native API, T1059.001:   PowerShell \\ 

Establish persistence &
  9 & create persistence & 
  T1059.003: Windows Command Shell, T1059.001: PowerShell \\ 

Create tasks/services &
  7 & create task, create service & 
  T1059.003: Windows Command Shell, T1059.001: PowerShell,   T1105: Ingress Tool Transfer \\ 

Harvest   credential &
  6 & access credential, dump credential &
  T1566.001: Spearphishing Attachment, T1059.001: PowerShell \\ 

Mimic legitimate   applications &
  6 & fake Windows update, fake Java update, fake Dropbox uploader, fake flash installer &
  T1547.001: Registry Run Keys \\ 

Disable system  functionalities &
  6 & disable recovery, disable security tools &
  T1059.003: Windows Command Shell, T1059.001: PowerShell \\ 

Manipulate   accounts &
  5 & create account, delete account &
  T1059.003: Windows Command Shell, T1106: Native API \\ 

Spread laterally &
  5 & lateral movement &
  T1059.001: PowerShell \\ 

Escalate   privilege &
  4 & privilege escalation &
  T1059.001: PowerShell, T1106: Native API \\ 

Hide command-and-control traffic &
  4 & obfuscate C2 traffic, obfuscate C2 location &
  T1071.004: DNS, T1071.001: Web Protocols \\ 
Query Registry &
  2 & read registry, access registry &
  T1082: System Information Discovery \\ \bottomrule
\end{tabular}%
\end{table}

\begin{table}[]
\centering
\scriptsize
\caption{Identified open codes in the \textbf{System Abuse} axial code}
\label{tab:axialSystemAbuse}
\begin{tabular}{p{3cm}rp{5cm}p{6cm}} \toprule
\textbf{Open codes} &
  \textbf{Count} &
  \textbf{Selected associated labels} &
  \textbf{Techniques} \\ \midrule

Shell interpreters &
  399 & cmd.exe, powershell, bash, visual basic interpreter, javascript interpreter, python interpreter, rundll32.exe &
  T1059.003: Windows   Command Shell, T1059.001: PowerShell \\ 

Network and data   communication protocols &
  388 & HTTP/S, FTP, SSH, TCP, DNS, POP, IMAP, websocket, wedav, SCP &
  T1071.004: DNS, T1041: Exfiltration Over C2 Channel, T1105:   Ingress Tool Transfer, T1071.001: Web Protocols \\ 

System API &
  237 & CreateProcess, GetProcAddress, LoadLibrary, ShellExecuteW, VirtualAlloc & 
  T1082: System Information Discovery, T1057: Process   Discovery, T1106: Native API, T1033: System Owner or User Discovery \\ 
  
Registry &
  171 & Windows Registry &
  T1059.001: PowerShell, T1082: System Information Discovery,   T1547.001: Registry Run Keys, T1059.003: Windows Command Shell, T1106: Native   API \\ 
  
Command line tools &
  155 & wget, curl, systeminfo, vssadmin, ver, fsutil, tasklist, psexec, ipconfig, whoami, certutil, ps, plist &
  T1033: System Owner or User Discovery, T1105: Ingress Tool   Transfer, T1082: System Information Discovery, T1057: Process Discovery,   T1071.001: Web Protocols, T1016: System Network Configuration Discovery \\ 
  
Windows Management Instrumentation &
  8 & Windows Management Instrumentation &
  T1059.001: PowerShell, T1033: System Owner or User   Discovery, T1082: System Information Discovery, T1059.003: Windows Command   Shell, T1057: Process Discovery \\ 
  
Third-party API &
  5 & Dropbox, ADFS, MS Exchange, MS Outlook &
  T1041: Exfiltration Over C2 Channel \\ 
  
Task scheduler/service &
  4 & task scheduler, windows service &
  T1059.001: PowerShell, T1070.004: File Deletion, T1105:   Ingress Tool Transfer \\ 
  
Account &
  3 & OS account, domain account &
  T1566.001: Spearphishing Attachment, T1204.002: Malicious   File \\ 
  
Security software whitelisting &
  1 & whitelisting feature in antivirus &
  T1059.001: PowerShell \\ 
  
Operating system authentication &
  1 & pass the hash &
  T1059.001: PowerShell \\ \bottomrule
\end{tabular}%
\end{table}

\begin{table}[]
\centering
\scriptsize
\caption{Identified open codes in the \textbf{Information} axial code}
\label{tab:axialInformation}
\begin{tabular}{p{3cm}rp{5cm}p{6cm}}
\toprule
\textbf{Open codes} &
  \textbf{Count} & \textbf{Selected associated labels} &
  \textbf{Techniques} \\ \midrule

System information &
  1055 & OS, CPU, memory, hardware, volumes, display, processor architecture, updates, patches, service packs, IP address, routing, proxy, MAC address, open ports &
  T1041: Exfiltration   Over C2 Channel, T1082: System Information Discovery, T1027: Obfuscated files   or information, T1106: Native API, T1057: Process Discovery, T1083: File and   Directory Discovery, T1016: System Network Configuration Discovery, T1005: Data from Local Systems \\ 
  
User   information &
  228 & user names, groups, privileges, permissions, root users, admin users, location, language, uptime, logged in status &
  T1082: System Information Discovery, T1033: System Owner or   User Discovery \\ 
  
System security information &
  22 & antivirus, firewall, presence of debugger and sandbox &
  T1082: System Information Discovery, T1106: Native API,   T1057: Process Discovery, T1083: File and Directory Discovery \\ 
  
Malware deployment status &
  9 & C2 beacon, whether payloads are received, current malware version, presence of rival malware &
  T1082: System Information Discovery, T1083: File and   Directory Discovery \\ \bottomrule
\end{tabular}%
\end{table}

\begin{table}[]
\centering
\scriptsize
\caption{Identified open codes in the \textbf{Malicious Object} axial code}
\label{tab:axialMaliciousObject}
\begin{tabular}{p{3cm}rp{5cm}p{6cm}} \toprule
\textbf{Open codes}                             & \textbf{Count} & \textbf{Selected associated labels} & \textbf{Techniques}                                                                                      \\ \midrule
Source code and scripts &
  315 & PowerShell script, batch script, bash script, vbscript, lua script, javascript &
  T1566.001:   Spearphishing Attachment, T1059.001: PowerShell, T1059.003: Windows Command   Shell, T1027: Obfuscated files or information, T1140: De-obfuscation \\ 
  
Secondary stage   malware                     & 160 & additional malware, malware updates, malware plugins, modular malware           & T1105: Ingress Tool Transfer                                                                             \\ 

Weaponized   documents &
  125 & office documents containing macro, payload hidden in image &
  T1566.001: Spearphishing Attachment, T1204.002: Malicious   File, T1140: De-obfuscation, T1027: Obfuscated files or information \\ 
  
Payload and C2   communication                & 122   & payload, shellcode, C2 commands, C2 beacon         & T1071.001: Web Protocols, T1140: De-obfuscation, T1027:   Obfuscated files or information                \\ 

Malware Footprints                            & 111 & tmp folder, appdata folder, malware logs, malware installers           & T1070.004: File Deletion                                                                                 \\ 

Malware   configurations                      & 58   & malware configuration files, malware configuration in registry, malware configuration in environment variable          & T1140: De-obfuscation, T1070.004: File Deletion, T1027:   Obfuscated files or information                \\ 

Persisting   malware &
  46 & startup folder, registry keys, keylogger, webshell, service, cryptominers, Onedrive, Outlook mailer, Dropbox &
  T1041: Exfiltration Over C2 Channel, T1059.001: PowerShell,   T1547.001: Registry Run Keys, T1105: Ingress Tool Transfer, T1071.001: Web   Protocols \\ 
  
Hidden macro, shellcode, and executables &
  26 & macro, scripts inside image, hidden content in document footer in white color, hidden content in spreadsheet formula & 
  T1566.001: Spearphishing Attachment, T1059.001: PowerShell,   T1059.003: Windows Command Shell, T1070.004: File Deletion, T1027: Obfuscated   files or information, T1140: De-obfuscation, T1204.002: Malicious File,   T1083: File and Directory Discovery \\ 
  
Obfuscated files                              & 24   & obfuscated code, obfuscated scripts, obfuscated payloads          & T1566.001: Spearphishing Attachment, T1140: De-obfuscation,   T1070.004: File Deletion                   \\ 

Fake installers                               & 24     & fake antivirus, fake flash installer, fake desktop tools        & T1566.001: Spearphishing Attachment, T1204.002: Malicious   File, T1027: Obfuscated files or information \\ 

Tools and Libraries                           & 22       & dll, .net assemblies, utility programs such as 7z      & T1140: De-obfuscation, T1027: Obfuscated files or   information                                          \\ 

Third party   information repositories & 9       & Github, Google Drive       & T1105: Ingress Tool Transfer, T1071.001: Web Protocols                                                   \\ 

Scheduled tasks/services                        & 6     & scheduled tasks and services running malicious programs         & T1059.003: Windows Command Shell, T1070.004: File Deletion                                               \\ 

C2 server                              & 6    & C2 server, adversary controlled webpage, compromised third party webpage          & T1071.004: DNS, T1071.001: Web Protocols                                                                 \\ 
 
Registry keys                          & 4      & malware configuration stored in registry keys        & T1070.004: File Deletion                                                                                 \\ 

Exploits                               & 4      & browser exploits, printer exploits, webserver exploits        & T1566.001: Spearphishing Attachment                                                                      \\ 

HTML                                   & 4        & HTA, HWP, CHM, HTML files      & T1566.001: Spearphishing Attachment                                                                      \\ 

Malware secrets                        & 4     & malware decryption keys, C2 server passwords         & T1140: De-obfuscation, T1027: Obfuscated files or   information                                          \\ 

Email   attachments                    & 3      & document attachment, spreadsheet attachments, image attachments, installer attachments        & T1027: Obfuscated files or information                                                                   \\ 

Shortcuts                              & 3     & .lnk, shortcut files         & T1566.001: Spearphishing Attachment, T1059.003: Windows   Command Shell                                  \\ 

Links                                  & 2      & URLs to download malware, C2 url, compromised website url        & T1566.001: Spearphishing Attachment                                                                      \\ 

First stage   downloader               & 2      & dropper, loader        & T1105: Ingress Tool Transfer       
\\ \bottomrule
\end{tabular}%
\end{table}

\begin{table}[]
\centering
\scriptsize
\caption{Identified open codes in the \textbf{Affected Object} axial code}
\label{tab:axialAffetedObject}
\begin{tabular}{p{3cm}rp{5cm}p{6cm}}
\toprule
\textbf{Open codes}      & \textbf{Count} & \textbf{Selected associated labels} & \textbf{Techniques}                                                    \\ \midrule
Registry keys          & 164   & registry run key, registry runonce key         & T1547.001: Registry Run   Keys, T1027: Obfuscated files or information \\ 

Sensitive documents, images, media files & 129 & documents, spreadsheets, powerpoints, pdfs, images, media files, home folder, desktop, recent files, organizational files & T1041: Exfiltration Over C2 Channel, T1082: System   Information Discovery, T1083: File and Directory Discovery, T1005: Data from Local Systems \\ 

System configuration & 72   & startup folder, kernel module & T1059.003: Windows Command Shell, T1547.001: Registry Run   Keys       \\ 

System files                             & 27 & backup, restore points, volume shadow copy, driver files, boot sector & T1140: De-obfuscation, T1083: File and Directory Discovery,   T1070.004: File Deletion                          \\ 

Process                & 24             & running process, critical system process, process memory, browser, banking applications, venv, MSBuild, CLR, cryptominers, ADFS server &  T1057: Process Discovery, T1106: Native API, T1059.001:   PowerShell, T1005: Data from Local Systems   \\ 

Credentials                              & 14 & LSASS credential cache, NTLM credential cache, domain credential cache, browser credentials, password hash  & T1566.001: Spearphishing Attachment, T1041: Exfiltration   Over C2 Channel, T1059.001: PowerShell, T1005: Data from Local Systems               \\ 

Log files              & 8      & debug messages, alert messages, system logs        & T1070.004: File Deletion                                               \\ 

Account                & 5     & OS account, domain account         & T1059.003: Windows Command Shell, T1106: Native API                    \\ 

Payment Information & 5 & payment card data, credit card data & T1005: Data from Local Systems \\ 

Software configuration & 3  & enabling macro by default            & T1059.001: PowerShell, T1547.001: Registry Run Keys, T1005: Data from Local Systems                  \\  

Local files            & 3         & local files     & T1070.004: File Deletion, T1027: Obfuscated files or   information     \\ 

Installed application  & 2 & browsers, MS exchange, MS outlook             & T1071.001: Web Protocols                                               \\ 

Private conversation data & 2 & text messages, instant message conversations & T1005: Data from Local Systems \\ 

Browser history & 1 & browsing history, cookies & T1005: Data from Local Systems \\ 
Local database & 1 & locally installed database & T1005: Data from Local Systems \\ 
\bottomrule
\end{tabular}%
\end{table}

\subsection{How the prevalent techniques are implemented}
\label{rq1:implement}
After performing Step 5, reported in Section~\ref{rq1:prevalentTechniques}, we obtain 73 open codes. The open codes are grouped into five axial codes, each representing specific characteristics regarding the implementations of prevalent techniques. We report the open codes of each axial code in Table 6 - 10. The count of an open code denotes the count of procedures labeled to the open code. Below, we report the axial codes, count, $\kappa$ agreement score and discuss our observations. The count of an axial code denotes the total counts of corresponding open codes. 

\subsubsection{\textbf{Action (n=1898) ($\kappa=0.89$):}}
\textit{Action refers to the specific activity performed in a technique.} For example, encrypting or encoding files are actions in the case of \textit{T1027: Obfuscated files or information} technique. We report the identified open codes for the \textit{Action} axial code in Table~\ref{tab:axialAction}. The top five open codes in the table account for 81\% of the labels in the \textit{Action} axial code. Adversaries apply obfuscation algorithms on malicious executables, payloads, and source code by encrypting, encoding, compressing, and applying source code obfuscation. Adversaries deploy malware to the victim system by downloading or transferring malicious scripts, payloads, and tools into the victim systems. Adversaries persist the malware by setting the \verb|run| registry key in Windows systems. Adversaries clean all the malware footprints after compromise and exfiltration by deleting all the malicious and related files and logs, uninstalling programs, and, thus, restoring the victim system configuration to the pre-compromise state. However, we also observe victim users unknowingly act on the attackers' behalf. Through phishing emails, adversaries trick users into opening documents and enable malicious and hidden macro/content. Thus, through user execution, attackers can successfully run arbitrary codes.

\subsubsection{\textbf{System abuse (n=1372) ($\kappa=0.94$):}}
\textit{System abuse refers to using system components, features, protocols, and APIs for malicious purposes.} For example, in the case of \textit{T1059.001: Powershell}, adversaries use the Powershell commands for malicious purposes, and thus, using Powershell is an abuse of the shell command - which is a system feature. We report the identified open codes for the \textit{System Abuse} axial code in Table~\ref{tab:axialSystemAbuse}. The top five open codes account for 98.4\% of the labels in the \textit{System abuse} axial code. The topmost abused system feature is the shell interpreters that come prebuilt with operating systems, such as Windows command shell, Unix shell, or others, such as Python interpreter. Adversaries use shell interpreters to execute arbitrary commands. The second most abused system feature is the network and data communication protocols such as HTTP, DNS, FTP, and SSH, through which adversaries establish C2, communicate remote commands, and transfer data from victims' end to adversaries' end. The third most abused system feature is the use of system APIs provided by the operating systems. For example, malware can create a new process in a Windows system by calling \verb|CreateProcess()| function from \verb|Win32.h| library, which comes prebuilt with Windows systems. Using these API functions benefits adversaries in persisting because anti-malware programs may not detect the API calls as part of a malicious operation. The fourth topmost abused feature is the use of Windows Registry or Windows startup folder to configure malware to run at system startup but at the background so users may not notice them. The fifth topmost abused system features are native or publicly available command-line tools such as \verb|systeminfo|, \verb|psexec|, \verb|curl|, \verb|wget| for deploying malware and performing internal reconnaissance into the victim system.

\subsubsection{\textbf{Information (n=1314) ($\kappa=0.86$):}}  
\textit{Information refers to the user and system-specific data that can aid adversaries in obtaining their malicious goals.} For example, in the case of \textit{T1082: System Information Discovery} technique, the operating system information can aid adversaries in customizing their malware for the specific version. We report the identified open codes for the \textit{Information} axial code in Table~\ref{tab:axialInformation}. The most collected information is (a) system information (OS, hardware, patches, processor, memory, disk), network information (IP, DNS, routing), currently running processes in a system, (b) user account metadata (privilege, permission, domain group, logged-in status), (c) presence of antivirus systems and other security measures, such as sandboxes and debugger tools, and (d) information related to command-and-control and malware installation. We also observe that, among all the open codes across the five axial codes, victim system information is the topmost open code in terms of label count. The observation suggests that adversaries may spend considerable effort in understanding the victim systems before starting malicious operations such as stealing data or encrypting data for ransom.

\subsubsection{\textbf{Malicious object (n=1080) ($\kappa=0.91$):}}
\textit{Malicious object refers to the files, scripts, executables, and applications used by adversaries.} For example, in the case of \textit{T1105: Ingress tool transfer} technique, the adversary imports required tools and scripts, which are malicious objects. We report the identified open codes for the \textit{Malicious Object} axial code in Table~\ref{tab:axialMaliciousObject}. The top five open codes account for 77.1\% of labels in the \textit{Malicious Object} axial code, which we discuss in the following. \textbf{Source code and scripts:} Adversaries deliver malicious scripts, which are directly executable from the command line, such as batch scripts, PowerShell scripts, visual basic scripts, bash scripts, etc. Adversaries oftentimes deliver custom source code of malware which they compile directly in the victim's system. \textbf{Second-stage malware:} APT attackers first collect system-specific information of a victim system, obtain or develop suitable malware and tools, and import them into victim systems. The information is collected by first-stage malware, delivered during the initial compromise. The first and second-stage malware persist in the system, collects system information, and exfiltrates data. \textbf{Weaponized documents:} Through phishing emails, adversaries send document files, scripts, and fake installers, which may contain malicious macro and executable contents. Fake installers use the icon of well-known desktop software. However, the installers deploy the initial malware loader to establish the C2 channel and collect system information. \textbf{Payload and C2 communication:} Once adversaries establish an initial foothold, the delivered malware sends an initial beacon to the command-and-control (C2) server to establish a persistent communication channel between C2 and the victim machine. Then the adversaries send commands from C2 and malicious payload to the victim system. \textbf{Malware footprint:} During infection, adversaries use malicious tools and scripts, creating temporary files and logs. After fulfilling the malicious objectives, adversaries delete these tools, scripts, and temporary files from the system.
    

While we do not enumerate the rest of the open codes from the table, we observe that \textbf{adversaries can turn any benign digital artifact into a malicious one}. We list a few cases in the following. Registry keys and environment variables can be used to store malware configurations, such as the IP address of the C2 server. Cloud applications such as Dropbox and Onedrive can be used to exfiltrate sensitive documents. Third-party information repositories such as Github, and Google Drive can be used to host malware. 
 
\subsubsection{\textbf{Affected objects (n=460) ($\kappa=0.94$):}}
\textit{Affected object refers to the system assets impacted by a technique.} For example, the modified registry key is an affected object in the case of \textit{T1547.001 Registry Run Keys} technique. We report the identified open codes for the \textit{Affected Object} axial code in Table~\ref{tab:axialAffetedObject}. The top three open codes account for 90.4\% of labels in the \textit{Affected Object} axial code: (a) the \verb|run| key of the registry, through which malware can persist, (b) the personal and organizational files such as documents, images, and media files which adversaries collect and exfiltrate, (c) the system configurations such as registry value, permission, privileges - which adversaries can modify (d) system files such as critical operating system files, drivers, restore points, boot sectors - which adversaries can corrupt and overwrite and (e) operating system and application processes - to which adversaries can inject or side-load malicious processes. 


\section{Findings from RQ2}
\label{sec:findings_rq2}
This section discusses the findings for \textbf{RQ2}. Section~\ref{rq2:pairs}, and~\ref{topcorrelated} discuss the obtained recurring pairs of techniques. Section~\ref{relations}-~\ref{platform} discuss how techniques are combined in the recurring pairs. Section~\ref{rq2:centrality} discusses the techniques appearing in a relatively higher number of recurring pairs. 

\subsection{Summary of the identified pairs}
\label{rq2:pairs}
After performing the Step-2 of Section~\ref{rq2:methods}, we identify 425 pairs of techniques where each pair contains an antecedent and a consequent technique\footnote{The list of all 425 pairs can be found in Section 3 of the supplementary file}. In each pair, the antecedent and the consequent are correlated with at least $0.20$ Pearson Phi correlation coefficient. The antecedent and the consequent of all pairs are also correlated with statistical significance - which is determined by the Chi-Squared test. In the majority of the pairs (321 out of 425), we observe that the antecedent and consequent techniques are weakly correlated. We identified 73 moderate, 29 strong, and two very strong correlations. We also compute the increase of the probability of a consequent given the antecedent is present for each pair and find that the median increase in the consequent probability is $5.3$ times. The observation indicates that despite a majority of the pairs being weakly correlated, one technique from the pair can significantly impact the occurrence of the other. Overall, the pairs suggest that the execution of techniques can further open the attack vector or broaden the attack surface for executing other techniques. 


\begin{table}[]
\centering
\scriptsize
\caption{Recurring pairs with strong and very strong correlation}
\label{tab:topCorrelaedTechniques}
\begin{tabular}{lrr}
\toprule
\textbf{Pairs}                                                                                         & \textbf{Phi} & \textbf{Support} \\ \midrule
(C1) T1204.001: Malicious Link (TA0002: Execution) $\land$ T1566.002: Spearphishing Link (TA0001: Initial Access)                     & 0.82         & 0.10             \\
(C2) T1204.002: Malicious File (TA0002: Execution) $\land$ T1566.001: Spearphishing Attachment (TA0001: Initial Access)               & 0.78         & 0.23             \\
(C3) T1489: Service Stop (TA0040: Impact) $\land$ T1490:   Inhibit System Recovery (TA0040: Impact)                                     & 0.66         & 0.03             \\
(C4) T1585.002: Email Accounts (Resource   development) $\land$ T1585.001: Social Media Accounts (TA0042: Resource Development) & 0.63         & 0.01             \\
(C5) T1003.004: LSA Secrets (Credential   access) $\land$ T1003.005: Cached Domain Credentials (TA0006: Credential Access)      & 0.62         & 0.01             \\
(C6) T1003.002: Security Account Manager   (TA0006: Credential Access) $\land$ T1003.004: LSA Secrets (TA0006: Credential Access)       & 0.61         & 0.01             \\
(C7) T1567.002: Exfiltration to Cloud Storage   (TA0010: Exfiltration) $\land$ T1583.006: Web Services (TA0042: Resource Development)    & 0.56         & 0.02             \\
(C8) T1486: Data Encrypted for Impact (TA0040: Impact)   $\land$ T1490: Inhibit System Recovery (TA0040: Impact)                        & 0.56         & 0.03             \\
(C9) T1485: Data Destruction (TA0040: Impact) $\land$   T1561.002: Disk Structure Wipe (TA0040: Impact)                                 & 0.55         & 0.02             \\
(C10) T1123: Audio Capture (TA0009: Collection) $\land$   T1125: Video Capture (TA0009: Collection)                                      & 0.53         & 0.03             \\
(C11) T1589: Gather Victim Identity Information   (TA0043: Reconnaissance) $\land$ T1589.002: Email Addresses (TA0043: Reconnaissance)   & 0.53         & 0.01             \\
(C12) T1594: Search Victim-Owned Websites   (TA0043: Reconnaissance) $\land$ T1585.002: Email Accounts (TA0043: Reconnaissance)          & 0.51         & 0.01             \\
(C13) T1486: Data Encrypted for Impact (TA0040: Impact)   $\land$ T1489: Service Stop (TA0040: Impact)                                   & 0.51         & 0.03             \\
(C14) T1543.004: Launch Daemon (TA0003: Persistence)   $\land$ T1543.001: Launch Agent (TA0003: Persistence)                             & 0.46         & 0.01             \\
(C15) T1033: System Owner/User Discovery   (TA0007: Discovery) $\land$ T1082: System Information Discovery (TA0007: Discovery)           & 0.46         & 0.21             \\
(C16) T1555: Credentials from Password Stores   (TA0006: Credential Access) $\land$ T1003.005: Cached Domain Credentials (Credential   access) & 0.45 & 0.01 \\
(C17) T1594: Search Victim-Owned Websites   (TA0043: Reconnaissance) $\land$ T1598.003: Spearphishing Link (TA0043: Reconnaissance)      & 0.44         & 0.01             \\
(C18) T1069.002: Domain Groups (TA0007: Discovery)   $\land$ T1087.002: Domain Account (TA0007: Discovery)                               & 0.43         & 0.01             \\
(C19) T1213: Data from Information Repositories   (TA0009: Collection) $\land$ T1003.003: NTDS (TA0006: Credential Access)               & 0.43         & 0.01             \\
(C20) T1059.005: Visual Basic (TA0002: Execution) $\land$   T1566.001: Spearphishing Attachment (TA0001: Initial Access)                 & 0.43         & 0.12             \\
(C21) T1014: Rootkit (TA0005: Defense Evasion) $\land$   T1574.006: Dynamic Linker Hijacking (TA0003: Persistence)                       & 0.43         & 0.01             \\
(C22) T1016: System Network Configuration   Discovery (TA0007: Discovery) $\land$ T1082: System Information Discovery (TA0007: Discovery)              & 0.42 & 0.23 \\
(C23) T1078: Valid Accounts (TA0005: Defense Evasion)   $\land$ T1133: External Remote Services (TA0003: Persistence)                    & 0.42         & 0.02             \\
(C24) T1561.002: Disk Structure Wipe (TA0040: Impact)   $\land$ T1529: System Shutdown/Reboot (TA0040: Impact)                           & 0.42         & 0.01             \\
(C25) T1020: Automated Exfiltration   (TA0010: Exfiltration) $\land$ T1119: Automated Collection (TA0009: Collection)                    & 0.41         & 0.02             \\
(C26) T1585.001: Social Media Accounts   (TA0042: Resource Development) $\land$ T1566.003: Spearphishing via Service (Initial   access)        & 0.41 & 0.01 \\
(C27) T1016: System Network Configuration   Discovery (TA0007: Discovery) $\land$ T1033: System Owner/User Discovery (TA0007: Discovery) & 0.41         & 0.15             \\
(C28) T1204.001: Malicious Link (TA0002: Execution)   $\land$ T1204.002: Malicious File (TA0002: Execution)                              & 0.41         & 0.09             \\
(C29) T1059.005: Visual Basic (TA0002: Execution) $\land$   T1204.002: Malicious File (TA0002: Execution)                                & 0.41         & 0.12             \\
(C30) T1056.001: Keylogging (TA0009: Collection) $\land$   T1113: Screen Capture (TA0009: Collection)                                    & 0.40         & 0.11             \\
(C31) T1082: System Information Discovery   (TA0007: Discovery) $\land$ T1083: File and Directory Discovery (TA0007: Discovery)          & 0.40         & 0.26             \\ \bottomrule
\end{tabular}%
\end{table}

\subsection{Top pairs by correlation strength}
\label{topcorrelated}
We report the two very strongly correlated and 29 strongly correlated pairs in Table~\ref{tab:topCorrelaedTechniques}. Each pair shows (a) the two techniques and their corresponding tactics inside the parenthesis; and (b) Pearson's Phi coefficient and support (i.e., percentage of CTI reports where the pair is found). We observe ten pairs reflecting adversaries' focus on exploiting social engineering vectors combined with victim user execution. The pairs are: C1, C2, C4, C11, C12, C17, C20, C26, C28, and C29. The adversaries obtain organizational emails and social media footprints of potential victim users to facilitate malware delivery campaigns. The adversaries deliver malware hidden in attachments or download links. Victim users explicitly execute or install the downloaded malware into their systems. We observe five pairs where adversaries collect information on victim system architecture, network architecture, user privileges, and file contents. The pairs are: C15, C18, C22, C27, and C31. After the initial infection, the delivered malware performs these techniques simultaneously. We observe five pairs primarily reflecting the final stage of ransomware attacks. The pairs are: C3, C8, C9, C13, and C24. The adversaries disable/delete the operating systems' backup and system restore facilities and corrupt the file-system architecture.

We observe four pairs reflecting the techniques related to espionage and data-breach activities. The pairs are C7, C10, C25, and C30. The pairs suggest four specific espionage activities - audio, video, computer screen, and typed inputs by victims. We also observe that adversaries set up cloud services (e.g., Dropbox, Google Drive) and program the malware to exfiltrate the captured information automatically using the background services. We observe four pairs where adversaries compromise credentials: C5, C6, C16, and C19. We observe adversaries use password dumping tools to extract credentials from the cache or process memory from the operating systems. Finally, we observe three pairs through which adversaries can remain in victim systems in a hidden manner. C14 implies the use of background services for automatically collecting information. C21 implies injecting malicious processes into legitimate applications, so antivirus fails to detect those malicious processes. C23 denotes adversaries use compromised credentials to log in to remote services with legitimate user accounts. 

\begin{table}[]
    \centering
    \scriptsize
    \caption{Summary of the relationship types among the techniques in pairs}
    \label{tab:ruleTypes}
    \begin{tabular}{lp{10cm}rl}
        \toprule

        \textbf{Type} & \textbf{Definition} & \textbf{Count} & \textbf{$\kappa$ score} \\ \midrule

        Same asset & Two techniques abuse or affect the same system asset & 176 & 0.93** \\
        Follow & Consequent technique follows the execution of antecedent technique & 163 & 0.88** \\
        Implementation overlap & Implementation of one technique can also fully or partially implement the other technique & 123 & 0.91** \\
        Happens together & Antecedent and consequent can happen concurrently at the same phase of an attack & 88 & 0.89** \\
        Require & The implementation of the first technique is required to implement the second technique & 55 & 0.89** \\
        Alternative & Two techniques can aid adversaries in achieving the same goal & 38 & 0.75*  \\
        Same platform & Two techniques are targeted for same platform & 22 & 0.94** \\ 
        \bottomrule
        \multicolumn{4}{c}{*: substantial agreement, **: almost perfect agreement}
    \end{tabular}
\end{table}

\subsection{Identified relationships among the recurring pairs of techniques}
\label{relations}
After performing qualitative coding, and inter-rater agreement, we identify seven relationships among the recurring pairs. Table~\ref{tab:ruleTypes} summarizes the identified relation types and the corresponding inter-rater agreement score. Each type denotes the relationship types between the antecedent and the consequent. The relationships are not orthogonal, which means two techniques can be related for multiple reasons. In the following sections, from Sections~\ref{same-asset} to~\ref{platform}, we discuss the top five techniques for both support and Pearson's Phi of each relation types. Support denotes the percentage of CTI reports where we observe the pair. Phi denotes the strength of the correlation between the antecedent, and the consequent techniques in pairs. We also discuss the top techniques by degree centrality score across all relations (see Step-4 in Section~\ref{rq2:methods}) in Section~\ref{rq2:centrality}.

\begin{table}[]
\centering
\scriptsize
\caption{Top five pairs for each of support and Pearson's Phi in the \textbf{Same Asset} relation. The pairs are sorted by support value. The common asset is shown in the Italics}
\label{tab:topfiveassetrelation}
\begin{tabular}{p{70mm}lp{70mm}}
\toprule
\textbf{Pair} &
  \textbf{Sup(Phi)} &
  \textbf{Explanation} \\ \midrule
(C1) T1027:   Obfuscated Files or Information (Defense Evasion) $\land$ T1105: Ingress Tool   Transfer (Command and Control) &
  0.30 (0.28) &
  \textit{Malicious files} are obfuscated and   imported \\
(C2) T1027: Obfuscated Files or   Information (Defense Evasion) $\land$ T1071.001: Web Protocols   (Command and Control) &
  0.25 (0.26) &
  \textit{Malicious files} are obfuscated and communicated over HTTPS   protocol \\
(C3) T1204.002: Malicious File   (Execution) $\land$ T1566.001: Spearphishing Attachment (Initial Access) &
  0.23 (0.78) &
  \textit{Malicious files} are sent as phishing attachments and opened by   victim users \\
(C4) T1027: Obfuscated Files or   Information (Defense Evasion) $\land$ T1140: Deobfuscate/Decode Files or   Information (Defense Evasion) &
  0.22 (0.36) &
  \textit{Malicious files} are first obfuscated before delivery and   deobfuscated before use \\
(C5) T1105: Ingress Tool Transfer   (Command and Control) $\land$ T1140: Deobfuscate/Decode Files or Information   (Defense Evasion) &
  0.21 (0.25) &
  \textit{Imported malware} are obfuscated, later they are deobfuscated   before use \\
(C6) T1204.001: Malicious Link   (Command and Control) $\land$ T1566.002: Spearphishing Link (Initial Access) &
  0.10 (0.82) &
  \textit{Malicious links} are sent as phishing emails and visited by victim   users \\
(C7) T1489: Service Stop (Impact)   $\land$ T1490: Inhibit System Recovery (Impact) &
  0.03 (0.66) &
  \textit{Critical system processes/services} such as backup services and system restore services are stopped to inhibit recovery \\
(C8) T1003.002: Security Account   Manager (Credential Access) $\land$ T1003.004: LSA Secrets (Credential   access) &
  0.01 (0.61) &
  Both can aid adversaries in dumping \textit{credentials} \\
(C9) T1003.004: LSA Secrets   (Credential Access) $\land$ T1003.005: Cached Domain Credentials (Credential   access) &
  0.01 (0.62) &
  Both can aid adversaries in dumping \textit{credentials} \\ \bottomrule
\end{tabular}
\end{table}

\subsection{Same asset relation}
\label{same-asset}
The \textit{same asset} relation denotes that two techniques abuse or affect the same system asset. The asset can be a digital artifact in a system, such as a Windows registry. For example, these two techniques: \textit{T1012: Query registry} and \textit{T1112: Modify registry} have the \textit{same asset} relation. Because both querying and modifying registry values abuse the Windows registry to identify system information and modify the system behavior. An asset can also be an abstract system component. For example, these techniques \textit{T1003.001: LSASS Memory} and \textit{T1552.001: Credentials In Files} have the \textit{same asset} relation. By using the first technique, adversaries can dump credentials from process memory. The second technique allows adversaries to obtain credentials insecurely stored in files. In both techniques, the adversary can obtain the same type of asset (i.e., credential), which they can later abuse. 

We report the top five pairs in the same asset relationship for both support and Phi in Table~\ref{tab:topfiveassetrelation}. We observe three pairs (C1, C2, C5) in which adversaries apply obfuscation while importing malicious tools using web traffic. We observe two pairs (C8, C9) where adversaries apply credential dumping techniques. We observe two pairs (C3, C6) where phishing attachment leads to malicious execution.  Overall, we observe four types of assets where the pairs are applied: malware, malicious link, critical system process, and credentials.

The top five techniques by degree centrality score ($\delta$) in the same asset relationship are \textit{T1074.001: Local Data Staging}, \textit{T1003.001: LSASS Memory},  \textit{T1041: Exfiltration Over C2 Channel}, \textit{T1560.001: Archive via Utility}, and \textit{T1133: External Remote Services}. \textit{T1074.001}, \textit{T1041}, and \textit{T1560.011} techniques relate to other techniques in the identified pairs by the same asset: collected files and data of interest among adversaries. \textit{T1003.001} and \textit{T1133} techniques relate to other techniques in the identified pairs by the same asset: collected credentials. 

The topmost technique in the $\delta$ score is the \textit{T1074.001: Local Data Staging} technique, where adversaries stage all the collected information in a single location before exfiltration. The technique is related to techniques of three tactics reported as follows. In all tactics, the common asset is the collected files and data from the victim system. \textbf{\textit{TA0009: Collection}}: \textit{T1074.001} is related to the following \textit{TA0009: Collection} techniques: \textit{T1005: Data from Local System}, \textit{T1025: Data from Removable Media}, \textit{T1560.001: Archive via Utility}, \textit{T1113: Screen Capture}, \textit{T1119: Automated Collection}, \textit{T1560.003: Archive via Custom Method}, \textit{T1074.002: Remote Data Staging}. Because adversaries collect data from victim systems using these techniques. Hence, the collected data is the asset, and adversaries stage the asset locally in the victim system. \textbf{\textit{TA0007: Discovery}}: \textit{T1074.001} relates to \textit{T1120: Peripheral Device Discovery}, \textit{T1083: File and Directory Discovery}, and \textit{T1217: Browser Bookmark Discovery} techniques. Because adversaries collect peripheral devices, file-system, and bookmark information from the local system and stage the information before exfiltration. \textbf{\textit{TA0010: Exfiltration}}: \textit{T1074.001} is related to \textit{T1041: Exfiltration Over C2 Channel} technique because adversaries first stage the data and then exfiltrate them later.

Techniques with a relatively higher $\delta$ in the same asset relationship imply a relatively high number of other techniques depend on the execution of these techniques through the same assets. Such as \textit{T1003.001: LSASS Memory} can enable adversaries to obtain credentials for remote logins, for example, \textit{T1021.001: Remote Desktop Protocol}. The $\delta$ may imply that many different techniques may exist to abuse the same asset category. For example, \textit{T1003.001: LSASS Memory} and \textit{T1555: Credentials from Password Stores} enable adversaries to obtain credentials. Observing the top five techniques, we identify that credentials, sensitive information, and system information are the assets that are the primary target of the adversaries where they combine the techniques most. 

Practitioners should prioritize mitigating these techniques because they offer different ways to open avenues for subsequent attack options. The same-asset relation can aid practitioners in the following ways. If two techniques work on the same asset, the asset can be protected so that further attacks can not happen. For example, suppose \textit{LSA} component of Windows systems can be secured from credential dumping tools, such as MimiKatz~\cite{S0002}. In that case, practitioners can block one way of obtaining credentials and perform remote authentication. Two techniques working on similar types of assets - may indicate different possible attack paths for a technique. For example, an adversary can obtain credentials from files, password managers, and process memory. Securing all these three sources of passwords narrows down the available options for an attacker to obtain a credential. 
\begin{table}[]
\centering
\scriptsize
\caption{Top five pairs for each of support and Pearson's Phi in the \textbf{Follow} relation. The pairs are sorted by support value. The second one follows the first technique in a pair}
\label{tab:toptenfollow}
\begin{tabular}{p{70mm}lp{70mm}}
\toprule
\textbf{Pair} &
  \textbf{Sup(Phi)} &
  \textbf{Explanation} \\ \midrule
(C1) T1082:   System Information Discovery (Discovery) $\land$ T1105: Ingress Tool Transfer   (Command and Control) &
  0.32 (0.33) &
  After identifying victim system   information, adversaries import customized malicious tools into victim systems \\
(C2) T1082: System Information   Discovery (Discovery) $\land$ T1027: Obfuscated Files or Information (Defense   evasion) &
  0.27 (0.26) &
  Adversaries deliver malware in an obfuscated state. The   malware later performs OS fingerprinting \\
(C3) T1082: System Information   Discovery (Discovery) $\land$ T1071.001: Web Protocols (Command and Control) &
  0.26 (0.26) &
  Adversaries deliver malware using web protocols. The malware   later performs OS fingerprinting \\
(C4) T1083: File and Directory   Discovery (Discovery) $\land$ T1105: Ingress Tool Transfer   (Command and Control) &
  0.23 (0.20) &
  After identifying the filesystem information of victim systems,   adversaries import customized malicious tools into victim systems \\
(C5) T1105: Ingress Tool Transfer   (Command and Control) $\land$ T1070.004: File Deletion (Defense Evasion) &
  0.22 (0.26) &
  Adversaries deliver malware into victim systems. After the   malicious objective, such as exfiltration, is fulfilled, adversaries delete   the malware footprint from the victim system \\
(C6) T1027: Obfuscated Files or   Information (Defense Evasion) $\land$ T1140: Deobfuscate/Decode Files or   Information (Defense Evasion) &
  0.22 (0.36) &
  Once adversaries obfuscate source code, file, or information   using encoding, encryption, or compression, then later, they have to decode,   decrypt, or decompress the files \\
(C7) T1566.001: Spearphishing   Attachment (Initial Access) $\land$ T1059.005: Visual Basic (Execution) &
  0.12 (0.43) &
  Adversaries send phishing attachments containing visual basic   scripts. If users open the email and click the attachment, the script will   be executed \\
(C8) T1012: Query Registry   (Discovery) $\land$ T1112: Modify Registry (Defense Evasion) &
  0.07 (0.36) &
  Adversaries first search specific key-value pairs in the   Registry and can modify the value of a given Registry key (e.g., run key) \\
(C9) T1594: Search Victim-Owned   Websites (Reconnaissance) $\land$ T1598.003: Spearphishing Link   (Reconnaissance) &
  0.01 (0.44) &
  Adversaries study the organizational websites of potential   victims for victim identity information. such as email addresses. Later adversaries send phishing emails to  those addresses \\
(C10) T1003.003: NTDS (Credential   access) $\land$ T1213: Data from Information Repositories (Collection) &
  0.01 (0.43) &
  Adversaries dump Windows domain credentials. Later, they use the   credentials for SharePoint servers and collect sensitive data \\ \bottomrule
\end{tabular}%
\end{table}

\subsection{Follow relation}
\label{follow}
The \textit{follow} relation between the two techniques denotes the execution of the first technique happens after the execution of the second technique. For example, \textit{T1027: Obfuscated Files or Information}, and \textit{T1140: Deobfuscate/Decode Files or Information} techniques have a \textit{follow} relationship. Because an adversary first obfuscated files, import the files to the victims' end, and then deobfuscate the files before use. We report the top five pairs in the \textit{follow} relationship for both support and Phi in Table~\ref{tab:toptenfollow}. We observe five pairs (C1, C2, C3, C4, C8) where adversaries first apply techniques of the \textit{TA0007: Discovery} tactic to understand the victim system. The understanding enables them to determine what tools and scripts suit the victim. The understanding aids adversaries in determining the next course of action, such as \textit{T1105}, where they download secondary-stage malware. We identify two pairs (C7 and C9) suggesting a chain of events: adversaries first identify social footprint and send phishing emails leading to malicious execution. Another important observation is that adversaries delete malware traces, system logs, temporary files, etc, which were brought to the victims' end through T1105, as suggested in C5. 

Unlike the \textit{asset} relation, the \textit{follow} relation is directional. For example, (\textit{T1027}, \textit{T1140}) makes sense, but (\textit{T1140}, \textit{T1027}) does not make sense because a file first is obfuscated and then deobfuscated to its original state, while the opposite is not true. Hence, we computed both in-degree ($\delta_i$) and out-degree centrality ($\delta_o$) score techniques connected by the \textit{follow} relation. 

The top five techniques in the $\delta_i$ in the \textit{follow} relationship are \textit{T1070.004: File Deletion}, \textit{T1041: Exfiltration Over C2 Channel}, \textit{T1082: System Information Discovery}, \textit{T1560.001: Archive via Utility}, \textit{T1071.001: Web Protocols}. The relatively higher $\delta_i$ suggests that these techniques are primarily used in a later stage of an APT attack. For example, attackers compromise a system, collect data, exfiltrate the data (i.e., T1041), and delete all the traces of malicious activity from the system (i.e., T1070.004). 

The topmost technique in the $\delta_i$ is \textit{T1070.004: File Deletion}. We identify eleven techniques that \textit{the T1070.004} technique follows because adversaries use \textit{T1070.004: File Deletion} after successful compromise and exfiltration. The eleven techniques are \textit{T1027: Obfuscated Files or Information}, \textit{T1059.003: Windows Command Shell}, \textit{T1113: Screen Capture}, \textit{T1074.001: Local Data Staging}, \textit{T1005: Data from Local System}, \textit{T1547.001: Registry Run Keys / Startup Folder}, \textit{T1033: System Owner/User Discovery}, \textit{T1057: Process Discovery}, \textit{T1082: System Information Discovery}, \textit{T1083: File and Directory Discovery}, and \textit{T1105: Ingress Tool Transfer}.  Note that these eleven techniques occur at the early stage of infections - through which adversaries run malicious programs at background/boot to collect sensitive information. After exfiltration, adversaries delete all the footprints associated with these techniques and the malware which performed these techniques.

The top five techniques in the $\delta_o$ in the \textit{follow} relationship are \textit{T1105: Ingress Tool Transfer}, \textit{T1046: Network Service Discovery}, \textit{T1071.001: Web Protocols}, \textit{T1082: System Information Discovery}, and \textit{T1140: Deobfuscate/Decode Files or Information}. The relatively higher $\delta_o$ suggests that these techniques are primarily used in the early stage of APT attacks. For example, malicious scripts and tools are first imported (i.e., \textit{T1105}) into the victim system in obfuscated state (i.e., \textit{T1140}) so that malware scanners cannot analyze their content. 

The topmost technique in the $\delta_o$ is \textit{T1105: Ingress Tool Transfer}, where adversaries deliver malware to victim systems. We identify nine techniques from six different tactics that follow \textit{T1105}, reported in the following. \textbf{\textit{TA0007: Discovery}}: the \textit{T1082: System Information Discovery} and \textit{T1083: File and Directory Discovery} techniques follow \textit{T1105} because adversaries collect victim system information and file content following the delivery of the malware. \textbf{\textit{TA0002: Execution}}: \textit{T1059.003: Windows Command Shell} follows T1105 \textit{because} the delivered malware uses the command shell from the Windows operating system to execute malicious commands. \textbf{TA0005: Defense Evasion}: \textit{T1140: Deobfuscate/Decode Files or Information} and \textit{T1070.004: File Deletion} follow T1105 because the delivered malware needs to be: (a) decoded/decrypted before use; and (b) deleted after installation or exfiltration. \textbf{\textit{TA0003: Persistence}}: \textit{T1547.001: Registry Run Keys / Startup Folder} follows \textit{T1105} because adversaries will set the delivered malware as an auto-start program/service during system boot. \textbf{\textit{TA0009: Collection}}: \textit{T1005: Data from Local System}, \textit{T1113: Screen Capture} follow \textit{T1105} because the delivered malware collects sensitive data from local systems. \textbf{\textit{TA0010: Exfiltration}}: \textit{T1041: Exfiltration Over C2 Channel} follows \textit{T1105} because the delivered malware establishes the C2 channel from the victim system and exfiltrates the collected data using the channel.

Techniques with relatively higher $\delta_i$ or $\delta_o$ in the \textit{follow} relationship are important for defenders. The primary reason is these techniques are common converging and diverging techniques. A relatively high number of other techniques converge into or diverge from these techniques. Detecting and mitigating the techniques with high $\delta_o$ can heavily disrupt attackers in the early stage of an APT attack. For example, completely blocking any download activity into a system leaves attackers no choice but to import any malware. On the other hand, techniques with high $\delta_i$ often reflect the final targets of attackers. As a result, defenders should deploy maximum protection to the assets targeted by attackers.

\begin{table}[]
\centering
\scriptsize
\caption{Top five pairs for each of support and Pearson's Phi in the \textbf{Implementation Overlap} relation. The pairs are sorted by support value.}
\label{tab:topfiveoverlap}
\begin{tabular}{p{70mm}lp{70mm}}
\toprule
\textbf{Pair} &
  \textbf{Sup(Phi)} &
  \textbf{Explanation} \\ \midrule
(C1) T1027:   Obfuscated Files or Information (Defense Evasion) $\land$ T1105: Ingress Tool   Transfer (Command and Control) &
  0.30 (0.28) &
  Obfuscation is a step in data transfer. Ingress tool transfer technique uses data transfer to deliver malware \\
(C2)  T1071.001: Web Protocols   (Command and Control) $\land$ T1105: Ingress Tool Transfer   (Command and Control) &
  0.30 (0.32) &
  Ingress tool transfer technique uses data transfer to deliver malware. Web protocols such as HTTPS are used in data transfer \\
(C3)  T1059.003: Windows Command   Shell (Execution) $\land$ T1105: Ingress Tool Transfer (Command and Control) &
  0.28 (0.32) &
  Adversaries use commands like \verb|wget|, \verb|curl| to import malicious files   into victim systems \\
(C4)  T1027: Obfuscated Files or   Information (Defense Evasion) $\land$ T1071.001: Web Protocols   (Command and Control) &
  0.25 (0.26) &
  Adversaries encrypt, encode, or compress the malicious file and   Command and Control packets during their communication with HTTPS traffic \\
(C5)  T1059.003: Windows Command   Shell (Execution) $\land$ T1082: System Information Discovery (Discovery) &
  0.24 (0.23) &
  Adversaries use commands like \verb|systeminfo| which aids them in operating   system fingerprinting \\
(C6)  T1204.002: Malicious File   (Execution) $\land$ T1566.001: Spearphishing Attachment (Initial Access) &
  0.23 (0.78) &
  The adversaries first create a malicious document file and send those to the victim as phishing emails \\
(C7)  T1059.005: Visual Basic   (Execution) $\land$ T1566.001: Spearphishing Attachment (Initial Access) &
  0.12 (0.43) &
 The phishing attachment contains hidden malicious scripts using visual basic \\
(C8)  T1489: Service Stop (Impact)   $\land$ T1490: Inhibit System Recovery (Impact) &
  0.03 (0.66) &
  Adversaries can kill or disable system recovery services such as   backup services of the operating system \\
(C9)  T1589: Gather Victim Identity   Information (Reconnaissance) $\land$ T1589.002: Email Addresses   (Reconnaissance) &
  0.01 (0.53) &
  Victim email addresses are one kind of identity information of   the victim \\
(C10)  T1014: Rootkit (Defense   evasion) $\land$ T1574.006: Dynamic Linker Hijacking (Persistence) &
  0.01 (0.43) &
  Adversaries modifies the \verb|/etc/ld.so.preload| to overwrite the kernel   API calls to inject malicious process \\ \bottomrule
\end{tabular}
\end{table}

\subsection{Implementation Overlap relation}
\label{overlap}
The \textit{implementation-overlap} relation between two techniques denotes that the first technique is partially or fully implemented by implementing second technique, or vice versa. For example, the following two techniques: \textit{T1082: System Information Discovery} and \textit{T1059.003: Windows Command Shell}, have an \textit{implementation overlap} relationship between them. In Windows, adversaries can execute \verb|systeminfo| commands in the terminal, where the command returns various types of information on operating systems, devices, memory information, etc. Hence, the implementation overlap occurs: by using a command shell, the adversary obtains system information. We report the top five pairs in the \textit{implementation overlap} relationship for both support and Phi in Table~\ref{tab:topfiveoverlap}. We observe the \textit{implementation overlap} among  application layer protocols, use of obfuscation, and malware delivery. (C1, C2, C4). We also observe overlapping techniques in social engineering attacks (C6, C7, C9). 

We identify the following techniques having the top five highest $\delta$ in the \textit{implementation overlap relation}. These techniques are \textit{T1082: System Information Discovery}, \textit{T1119: Automated Collection}, \textit{T1083: File and Directory Discovery}, \textit{T1059.003: Windows Command Shell}, and \textit{T1047: Windows Management Instrumentation}. Implementing \textit{T1082}, \textit{T1119}, and \textit{T1083} techniques overlap with other techniques in collecting various properties of the victim system. On the other hand, adversaries use \textit{T1059.003} and \textit{T1047} techniques as low-level system feature to implement techniques with relatively advanced purposes. For example, the Windows command shell can execute arbitrary code, Windows management instrumentation facilitates remote login, scheduled execution, etc. 
    
The topmost technique in the $\delta$ is \textit{T1082: System Information Discovery}, which relates to nine other techniques across three tactics. We report our observations below for each of the three tactics. \textbf{\textit{TA0007: Discovery}:} \textit{T1082} overlaps with the following techniques: \textit{T1012: Query Registry}, \textit{T1016: System Network Configuration Discovery}, \textit{T1033: System Owner/User Discovery}, \textit{T1057: Process Discovery}, \textit{T1083: File and Directory Discovery}, \textit{T1518.001: Security Software Discovery}, \textit{T1124: System Time Discovery}. By using the technique, adversaries can obtain hardware, software, networking, and other relevant information about a victim system. These six \textit{TA0007: Discovery} techniques also implement system information discovery inherently. For example, identifying network information is also a part of identifying system information. Windows registry often stores Windows OS-specific information; thus, querying registry also implements \textit{T1082}. \textbf{\textit{TA0002: Execution}:} \textit{T1082} overlaps with \textit{T1059.003: Windows Command Shell} because operating systems provide shell commands, and adversaries can run the shell commands to identify system information. \textbf{\textit{TA0009: Collection}:} \textit{T1082} overlaps with \textit{T1113: Screen Capture} technique because adversaries often take screenshots of operating system diagnostic views (such as \verb|dxdiag.exe|) - which partially enables adversaries to obtain system information from the screenshot contents. 

Techniques with a relatively higher $\delta$ score in the \textit{implementation overlap} imply that a relatively high number of other techniques share the same implementation mechanism. For example, the \textit{T1059.003: Windows Command Shell} technique could be used by an attacker for a plethora of purposes ranging from downloading malware to executing malicious code. Practitioners should prioritize mitigating these techniques because they all use the same attack path leading to different attack scenarios. For example, if \textit{T1059.003: Windows Command Shell} can be protected from abuse, overlapping techniques, such as \textit{T1082: System Information Discovery}, can also be mitigated. 

\begin{table}[]
\centering
\scriptsize
\caption{Top five pairs for each of support and Pearson's Phi in the \textbf{Happen Together} relation. The pairs are sorted by support value.}
\label{tab:topFiveHappensTogether}
\begin{tabular}{p{65mm}lp{75mm}}
\toprule
\textbf{Pair} &
  \textbf{Sup(Phi)} &
  \textbf{Explanation} \\ \midrule
(C1) T1082:   System Information Discovery (Discovery) $\land$ T1083: File and Directory   Discovery (Discovery) &
  0.26 (0.40) &
  After making an initial   compromise, adversaries use these two techniques to collect information on the   victim system \\
(C2) T1057: Process Discovery   (Discovery) $\land$ T1082: System Information Discovery (Discovery) &
  0.23 (0.37) &
  After making an initial compromise, adversaries use these two   techniques to collect information on the victim system \\
(C3) T1016: System Network   Configuration Discovery (Discovery) $\land$ T1082: System Information Discovery (Discovery) &
  0.23 (0.42) &
  After making an initial compromise, adversaries use these two   techniques to collect information on the victim system \\
(C4) T1033: System Owner/User   Discovery (Discovery) $\land$ T1082: System Information Discovery (Discovery) &
  0.21 (0.46) &
  After making an initial compromise, adversaries use these two   techniques to collect information on the victim system \\
(C5) T1057: Process Discovery   (Discovery) $\land$ T1083: File and Directory Discovery (Discovery) &
  0.19 (0.38) &
  After making an initial compromise, adversaries use these two   techniques to collect information on the victim system \\
(C6) T1486: Data Encrypted for   Impact (Impact) $\land$ T1490: Inhibit System Recovery (Impact) &
  0.03 (0.56) &
  Adversaries disable all system recovery functions while they   encrypt the data of victims in a ransomware attack \\
(C9) T1486: Data Encrypted for   Impact (Impact) $\land$ T1489: Service Stop (Impact) &
  0.03 (0.51) &
  Adversaries kill critical system services while they encrypt   the data of victims in a ransomware attack \\
(C10) T1123: Audio Capture   (Collection) $\land$ T1125: Video Capture (Collection) &
  0.03 (0.53) &
  Adversaries capture the audio and video stream of a victim   user's computer to collect sensitive information \\
T1585.002: Email Accounts   (Resource Development) $\land$ T1585.001: Social Media Accounts (Resource   development) &
  0.01 (0.63) &
  Adversaries establish fake accounts to mimic legitimate   persons or organizations before phishing campaigns \\
T1594: Search Victim-Owned   Websites (Reconnaissance) $\land$ T1585.002: Email Accounts (Resource   development) &
  0.01 (0.51) &
  Adversaries investigate the website of potential victims, and establish fake accounts to mimic legitimate persons or organizations before   phishing campaigns \\ \bottomrule
\end{tabular}
\end{table}

\subsection{Happen together}
\label{happen}
The \textit{happen together} relation between two techniques denotes that adversaries may implement both techniques at a same phase of an APT attack. For example, these two techniques happen together: \textit{T1123: Audio Capture}, and \textit{T1125: Video Capture}. Because the attackers primarily use these two techniques while collecting audio/video-related information about victim users. We report the top five pairs in the \textit{happen together} relationship for both support and Phi in Table~\ref{tab:topFiveHappensTogether}. We observe techniques from the \textit{TA0007: Discovery} tactic occur together (C1-C5). 

The top five techniques in the $\delta$ in the \textit{happen together} relationship are \textit{T1016: System Network Configuration Discovery}, \textit{T1113: Screen Capture}, \textit{T1087.001: Local Account}, \textit{T1083: File and Directory Discovery}, and \textit{T1049: System Network Connections Discovery}. These techniques all aid adversaries in internal reconnaissance. Attackers need to implement these techniques before they start eventual malicious operations. Hence figuring out the internal details of a target system is crucial for attackers. Defenders have few options for mitigating these techniques. Operating systems provide native APIs for obtaining basic system information and telemetries. Blocking the execution of these APIs may hamper other software and critical system processes. The topmost technique in the $\delta$ is \textit{T1016: System Network Configuration Discovery}. We identify adversaries using the following ten other techniques simultaneously with \textit{T1016}, and all ten techniques are from the \textit{TA0007: Discovery} tactic. These ten techniques are \textit{T1007: System Service Discovery}, \textit{T1012: Query Registry}, \textit{T1033: System Owner/User Discovery}, \textit{T1087.001: Local Account}, \textit{T1082: System Information Discovery}, \textit{T1083: File and Directory Discovery}, \textit{T1049: System Network Connections Discovery}, \textit{T1057: Process Discovery}, \textit{T1069.001: Local Groups}, and \textit{T1124: System Time Discovery}. All techniques happening together also aid adversaries in discovering various details of a target system. In the technique, adversaries identify network-related information to determine how they can laterally move internally inside a victim organization. Happening together with the highest number of other TA0007: Discovery techniques suggests that lateral movement is one of the primary targets of adversaries. 

The techniques with relatively high $\delta$ in the \textit{happen together} relation denote that many other techniques occur at the same phase of an APT attack. Defenders can benefit from the situation as attackers implement these techniques at the same stage. APT attacks can be very challenging to detect from system alerts. However, the techniques in the \textit{happen together} relations may aid defenders in tracing the indicators of these techniques. For example, if defenders identify that a process is executing native APIs for operating system information, network information, and process information, defenders should investigate the situation further.

\begin{table}[]
\centering
\scriptsize
\caption{Top five pairs for each of support and Pearson's Phi in the \textbf{Require} relation. The pairs are sorted by support value. The pair denotes that the second technique requires the first technique.}
\label{tab:topFiveRequire}
\begin{tabular}{p{65mm}lp{75mm}}
\toprule
\textbf{Pair} &
  \textbf{Sup(Phi)} &
  \textbf{Explanation} \\ \midrule
(C1) T1566.001:   Spearphishing Attachment (Initial Access) $\land$ T1204.002: Malicious File   (Execution) &
  0.23 (0.78) &
  Adversaries send phishing   attachments. The attachment provides malicious executables hidden in the   attachment. \\
(C2) T1566.002: Spearphishing Link   (Initial Access) $\land$ T1204.001: Malicious Link (Execution) &
  0.10 (0.82) &
  Adversaries send phishing links. The links lead to compromised   websites that can exploit the browser a victim is using \\
(C3) T1566.001: Spearphishing   Attachment (Initial Access) $\land$ T1204.001: Malicious Link (Execution) &
  0.08 (0.35) &
  Adversaries send phishing attachments. The attachments contain   download links for malware \\
(C4) T1566.002: Spearphishing Link   (Initial Access) $\land$ T1204.002: Malicious File (Execution) &
  0.08 (0.34) &
  Adversaries send phishing links. The links provide download   links for malware that victim users would download if the users click the links \\
(C5) T1083: File and Directory   Discovery (Discovery) $\land$ T1025: Data from Removable Media (Collection) &
  0.03 (0.22) &
  Adversaries search for partitions representing removable   media. Later, they collect data from the removable media \\
(C6) T1078: Valid Accounts (Defense   evasion) $\land$ T1133: External Remote Services (Persistence) &
  0.02 (0.42) &
  Adversaries identify valid user accounts and compromise/obtain   credentials for the account. Later, they use the accounts for logging into   external remote services such as VPN, RDP, etc. \\
(C7) T1583.006: Web Services   (Resource Development) $\land$ T1567.002: Exfiltration to Cloud Storage   (Exfiltration) &
  0.02 (0.56) &
  Adversaries set up web services in cloud service providers,   which they would later use for exfiltrating sensitive data \\
(C8) T1583.001: Domains (Resource   development) $\land$ T1608.001: Upload Malware (Resource Development) &
  0.01 (0.37) &
  Adversaries first set up domains they will use for   command-and-control channels, delivering malware. Later, they   stage/upload/host malware into the domain/website. \\ \bottomrule
\end{tabular}
\end{table}

\subsection{Require}
\label{require}
The \textit{require} relation between two techniques denotes the execution of a technique requires the execution of another technique. For example, the \textit{T1133: External Remote Services} and \textit{T1078: Valid Accounts} techniques have a \textit{require} relationship. To log in to a remote service, attackers need a valid account. Like the \textit{follow} relation, \textit{require} relation is also directional. We report the top five pairs in the \textit{require} relationship for both support and Phi in Table~\ref{tab:topFiveRequire}. We observe five pairs (C1-C4, C8), each denoting a chain of dependencies. For example, C8 suggests adversaries set up domains (T1583.001) which facilitates hosting the malware (T1608.001) to be delivered to the victims' end. C6 suggests, compromising valid accounts can aid adversaries persists through logging into RDP mimicking a legitimate user. C7 suggest that adversaries sets up web services which later enable them to exfiltrate sensitive data via these web services, such as Dropbox. 

The following techniques have the top five $\delta_i$ scores in the \textit{require} relation: \textit{T1133: External Remote Services}, \textit{T1204.001: Malicious Link}, \textit{T1598.003: Spearphishing Link}, \textit{T1021.001: Remote Desktop Protocol}, \textit{T1021.002: SMB/Windows Admin Shares}. Techniques with a relatively high $\delta_i$ score denote requiring at least one among a relatively high number of other techniques to be implemented first. For example, attackers can launch phishing campaigns if they obtain a list of email addresses from either official websites, social media accounts, or a compromised mailbox. The topmost technique in the $\delta_i$ is \textit{T1133: External Remote Services}. The technique requires the implementation of at least one of the following eight techniques: \textit{T1078: Valid Accounts}, \textit{T1078.004: Cloud Accounts}, \textit{T1114.002: Remote Email Collection}, \textit{T1003.001: LSASS Memory}, \textit{T1003.002: Security Account Manager}, \textit{T1552.004: Private Keys}, \textit{T1003.003: NTDS}, \textit{T1589.001: Credentials}. We observe that the eight techniques enable adversaries to obtain compromised credentials for logging into external remote services with a valid account. 

The following techniques have the top five $\delta_o$ in the \textit{require} relation: \textit{T1003.001: LSASS Memory}, \textit{T1078: Valid Accounts}, \textit{T1583.001: Domains}, \textit{T1585.002: Email Accounts}, \textit{T1585.001: Social Media Accounts}. Techniques with a relatively high $\delta_o$ denote that at least one among a relatively higher number of other techniques requires the techniques to be implemented first. All these techniques are related to obtaining or identifying legitimate account information and credentials. One key takeaway from these top five is username or email accounts, or any account or identity-related information should be given similar attention to credentials. Publicly available email addresses and usernames of a system can also facilitate attackers to run phishing campaigns, internal phishing campaigns, or social engineering. The topmost technique in $\delta_o$ is \textit{T1003.001: LSASS Memory}, and the technique aids adversaries in obtaining credentials by dumping process memory. We find that adversaries require \textit{T1003.001: LSASS Memory} technique to implement the five following techniques: \textbf{\textit{TA0008: Lateral Movement:}} Adversaries use the dumped credential obtained from the \textit{T1003.001} technique to log into connected systems using RDP protocols (\textit{T1021.001: Remote Desktop Protocol}) and to share malicious files using Windows Shared Message Bus (\textit{T1021.002: SMB/Windows Admin Shares}). \textbf{\textit{TA0003: Persistence} and \textit{TA0005: Defense Evasion}:} Adversaries use the dumped credential obtained from the \textit{T1003.001} technique to log into remote systems using \textit{1133: External Remote Services}, and persist the connection with legitimate accounts (\textit{T1078: Valid Accounts}) without raising suspicions. \textbf{\textit{TA0009: Collection}:} Adversaries use the dumped credential obtained from the \textit{T1003.001} technique to log into desktop email applications to read the mailboxes (\textit{T1114.002: Remote Email Collection}).
    
From the defenders' point of view, techniques related by the \textit{require} relation inform the dependency chain among adversarial techniques. Techniques having a \textit{require} relation with many other techniques can be successfully implemented if only one requiring technique is implemented. Defenders, should mitigate all possible dependencies to mitigate these techniques. Similarly, techniques that are common dependencies among many other techniques suggest that mitigation of these techniques can automatically nullify the attack vector for dependent techniques. 

\begin{table}[]
\centering
\scriptsize
\caption{Top five pairs for each of support and Pearson's Phi in the \textbf{Alternative} relation. The pairs are sorted by support value.}
\label{tab:topFiveAlternative}
\begin{tabular}{p{65mm}lp{75mm}}
\toprule
\textbf{Pair} &
  \textbf{Sup(Phi)} &
  \textbf{Explanation} \\ \midrule
(C1) T1059.001:   PowerShell (Execution) $\land$ T1059.005: Visual Basic (Execution) &
  0.09 (0.28) &
  Both Powershell and VBScript   can execute malicious code in victim systems \\
(C2)  T1566.001: Spearphishing   Attachment (Initial Access) $\land$ T1566.002: Spearphishing Link (Initial   access) &
  0.08 (0.37) &
  Phishing links and attachments can both be used to deliver   malware. The links lead to downloading/installation of malware. The   attachment leads to the execution of malware. \\
(C3)  T1059.005: Visual Basic   (Execution) $\land$ T1059.007: JavaScript (Execution) &
  0.04 (0.29) &
  Both Javascript and VBScript can execute malicious code in   victim systems \\
(C4)  T1547.001: Registry Run Keys /   Startup Folder (Persistence) $\land$ T1547.009: Shortcut Modification   (Persistence) &
  0.04 (0.27) &
  Adversaries can modify the registry to set a program to run while the system is booting. Adversaries can modify the symbolic link to point to a malicious program that can   also lead to running the program while booting \\
(C5)  T1003.002: Security Account   Manager (Credential Access) $\land$ T1003.004: LSA Secrets (Credential   access) &
  0.01 (0.61) &
  Both techniques aid adversaries in dumping credentials \\
(C6)  T1585.002: Email Accounts   (Resource Development) $\land$ T1585.001: Social Media Accounts (Resource   development) &
  0.01 (0.63) &
  Both accounts can aid adversaries in mimicking legitimate   persons or organizations \\
(C7)  T1003.004: LSA Secrets   (Credential Access) $\land$ T1003.005: Cached Domain Credentials (Credential   access) &
  0.01 (0.62) &
  Both techniques aid adversaries in dumping credentials \\
(C8)  T1543.004: Launch Daemon   (Persistence) $\land$ T1543.001: Launch Agent (Persistence) &
  0.01 (0.46) &
  Both techniques aid adversaries in launching malicious programs   at startup in UNIX-based operating systems \\
(C9)  T1555: Credentials from   Password Stores (Credential Access) $\land$ T1003.005: Cached Domain   Credentials (Credential Access) &
  0.01 (0.45) &
  Both techniques aid adversaries in dumping credentials \\
(C10)  T1555: Credentials from   Password Stores (Credential Access) $\land$ T1555.003: Credentials from Web Browsers (Credential Access) &
  0.03 (0.40) &
  Both techniques aid adversaries in obtaining credentials of   web applications users sign in. \\ \bottomrule
\end{tabular}
\end{table}

\subsection{Alternative}
\label{alernative}
The \textit{alternative} relation between the two techniques implies the two techniques can be used as alternatives to one another. For example, \textit{T1555: Credentials from Password Stores}, and \textit{T1552.001: Credentials In Files} can be used as an alternative to one another. This relation implies the potential alternative paths taken by adversaries. After computing the $\delta$, the top five techniques in this relation are \textit{T1003.001: LSASS Memory}, \textit{T1003.004: LSA Secrets}, \textit{T1003.005: Cached Domain Credentials}, \textit{T1555: Credentials from Password Stores}, \textit{T1552.001: Credentials In Files}. We observe that all these techniques are related to obtaining credentials by different means. Meanwhile, three of these five techniques (i.e., \textit{T1003.001}, \textit{T1003.004}, \textit{T1003.005}) are specifically targeted for the Windows platform. We report the top five pairs in the alternative relationship for both support and Phi in Table~\ref{tab:topFiveAlternative}. We observe that the highest number of alternative techniques exist for obtaining credentials (C5, C7, C9, C10). Windows and Unix-based platforms have multiple alternatives for running malicious tasks in the background so that users do not notice them (C4, C8).  The topmost technique in the $\delta$: \textit{T1003.001: LSASS Memory}, relates to five techniques from the \textit{TA0006: Credential Access} tactic. These five techniques are: \textit{T1003.004: LSA Secrets}, \textit{T1003.005: Cached Domain Credentials}, \textit{T1555: Credentials from Password Stores}, \textit{T1552.001: Credentials In Files} and \textit{T1003.001: NTDS}. We observe two open-source password extraction tools for obtaining credentials using the above-mentioned techniques, such as Mimikatz~\cite{S0002} and Lazagne~\cite{S0349}. The techniques with relatively high $\delta$ in the \textit{alternative} relation denote that adversaries can use a relatively higher number of alternatives for the techniques. For example, adversaries can obtain credentials using the \textit{T1003.001: LSASS Memory technique}. However, five alternative techniques exist for T1003.001: \textit{T1003.004: LSA Secrets}, \textit{T1555: Credentials from Password Stores}, \textit{T1003.005: Cached Domain Credentials}, \textit{T1552.001: Credentials In Files}, \textit{T1003.003: NTDS}.  From the defender's point of view, techniques and their alternatives must be mitigated simultaneously to narrow down the attacker's options at a given stage. 

\begin{table}[]
\centering
\scriptsize
\caption{Top five pairs for each of support and Pearson's Phi in the \textbf{Platform} relation. The pairs are sorted by support value. The explanation column specifies the common platform type for both techniques in a pair}
\label{tab:topFivePlatform}
\begin{tabular}{lll}
\toprule
\textbf{Pair}                                                                              & \textbf{Sup(Phi)} & \textbf{Explanation}       \\ \midrule
(C1) T1059.003:   Windows Command Shell (Execution) $\land$ T1547.001: Registry Run Keys /   Startup Folder (Persistence) & 0.18 (0.25) & Windows \\
(C2) T1012: Query Registry   (Discovery) $\land$ T1112: Modify Registry (Defense Evasion)              & 0.07         (0.36)         & Windows \\
(C3) T1047: Windows Management   Instrumentation (Execution) $\land$ T1059.001: PowerShell (Execution) & 0.07         (0.24)         & Windows \\
(C4) T1047: Windows Management   Instrumentation (Execution) $\land$ T1053.005: Scheduled Task (Execution)                & 0.06 (0.26) & Windows \\
(C5) T1112: Modify Registry   (Defense Evasion) $\land$ T1543.003: Windows Service (Persistence)       & 0.06         (0.24)         & Windows \\
(C6) T1003.001: LSASS Memory   (Credential Access) $\land$ T1003.004: LSA Secrets (Credential Access)  & 0.01         (0.36)         & Windows \\
(C7) T1003.002: Security Account   Manager (Credential Access) $\land$ T1003.004: LSA Secrets (Credential   access)       & 0.01 (0.61) & Windows \\
(C8) T1543.004: Launch Daemon   (Persistence) $\land$ T1059.004: Unix Shell (Execution)                & 0.01         (0.36)         & Unix   \\
(C9) T1003.004: LSA Secrets   (Credential Access) $\land$ T1003.005: Cached Domain Credentials (Credential   access)      & 0.01 (0.62) & Windows \\
(C10) T1014: Rootkit (Defense   evasion) $\land$ T1574.006: Dynamic Linker Hijacking (Persistence)      & 0.01         (0.43)         & Unix   \\ \bottomrule
\end{tabular}
\end{table}

\subsection{Platform}
\label{platform}
The \textit{platform} relation between the two techniques implies that the two techniques are targeted for the same platform. For example, \textit{T1059.004: Unix Shell} and \textit{T1053.003: Cron} are both targeted for Unix platforms. For this relation, the top five techniques in $\delta$ are \textit{T1059.004: Unix Shell}, \textit{T1003.001: LSASS Memory}, \textit{T1003.004: LSA Secrets}, \textit{T1112: Modify Registry}, \textit{T1059.001: PowerShell}. Among the five, we observe one for the Unix platform, and the rest of the four techniques are targeted for the Windows platform. From the defender's point of view, the relation would enable them to detect indicators of techniques based on the specific platforms. The topmost technique in the $\delta$, \textit{T1059.004}, relates to other Unix platform-specific techniques: \textit{T1222.002: Linux and Mac File and Directory Permissions Modification}, \textit{T1053.003: Cron}, \textit{T1543.004: Launch Daemon}, \textit{T1543.001: Launch Agent}. Note that all four techniques can be implemented using the shell functionality built-in with the Unix operating systems. We report the top five pairs in the \textit{platform} relationship for both support and Phi in Table~\ref{tab:topFivePlatform}. We observe adversaries combine the most platform-specific techniques for Windows (eight out of ten). Another important takeaway is that we identify three pairs for obtaining credentials for Windows platforms (C6, C7, C9). The observation suggests that credentials stored in Windows OS are under a relatively greater threat than other platforms.  

\begin{figure*}
    \centering
    \includegraphics[width=\textwidth]{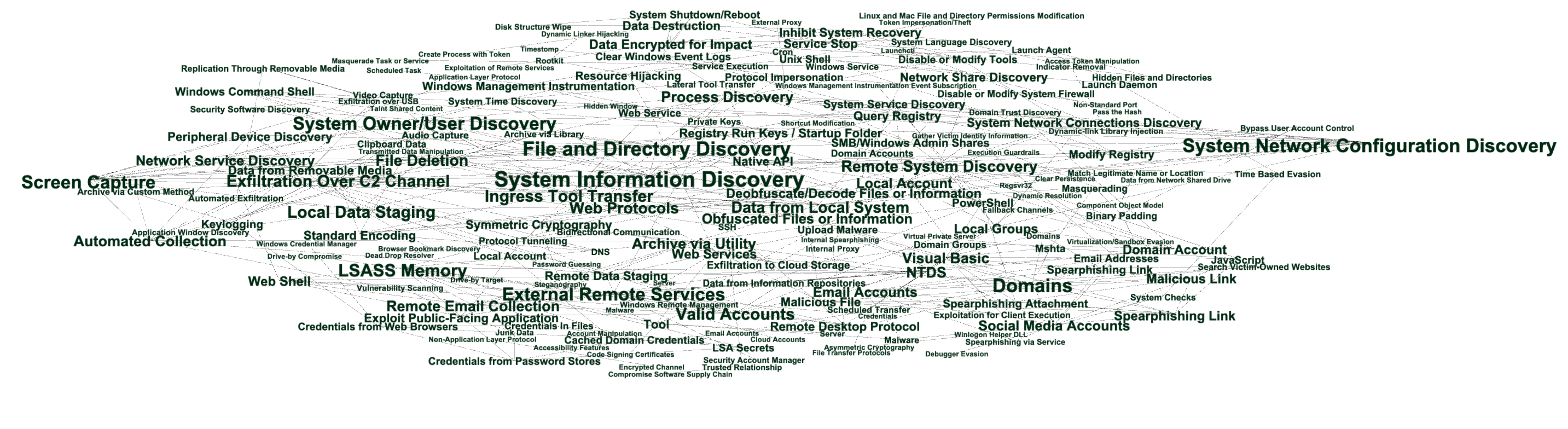}
    \caption{The graph showing all the techniques appeared in the 425 pair }
    \label{fig:centrality}
\end{figure*}

\subsection{Techniques that appeared in the highest number of pairs}
\label{rq2:centrality}
Fig.~\ref{fig:centrality} shows all techniques identified in 425 pairs in a graph data structure format, where nodes denote techniques, and edges denote two techniques appearing in the same pair. We compute the number ($\eta$) of other techniques combined with the technique for each. In the figure, technique names with bigger fonts denote a relatively higher value of $\eta$. The top ten techniques by $\eta$ are \textit{T1082: System Information Discovery} ($\eta=18$), \textit{T1083: File and Directory Discovery} ($\eta=16$), \textit{T1016: System Network Configuration Discovery} ($\eta=15$), \textit{T1583.001: Domains} ($\eta=15$), \textit{T1113: Screen Capture} ($\eta=14$), \textit{T1033: System Owner/User Discovery} ($\eta=14$), \textit{T1133: External Remote Services} ($\eta=14$), \textit{T1003.001: LSASS Memory} ($\eta=13$), \textit{T1074.001: Local Data Staging} ($\eta=12$), \textit{T1078: Valid Accounts} ($\eta=12$). A relatively higher value of $\eta$ suggests that the technique has a relatively higher utility for adversaries. Because the technique combines different types of attack vectors and broadens the attack surface. Similarly, from the defenders' point of view, these techniques are common weak points attack can come through. We observe three techniques from the \textit{TA0007: Discovery} tactic on the top in terms of $\eta$, indicating that victim system environment information has the highest utility for the attackers because the second-stage malware and tools depend on the information. We observe that the malicious domains adversaries set up for targeted victims are connected to many techniques, primarily social engineering and browser exploitation. We observe three techniques aiming to compromise credentials. And later, adversaries use the credentials to log in to remote services like RDP, VNC, etc., for mimicking legitimate users.

\section{Evaluation}
\label{sec:evaluation}
This section discusses the evaluation of our findings in RQ1 and RQ2. 

In RQ1, we identify a set of prevalent adversarial techniques. In RQ2, we identify a set of recurring pairs of adversarial techniques. To evaluate our findings, we first choose a set of the latest CTI reports that are not part of our dataset – we refer to the set of CTI reports as the set of \textit{unseen CTI reports}. We perform the following steps to construct a set of unseen CTI reports. \textbf{Step-1:} We find the date ($\Delta$) of the most recently published CTI reports from the dataset. \textbf{Step-2:} In Section~\ref{sec:method_dataset_summary}, we randomly selected 161 (25\%) CTI reports from our final dataset. From the 161 reports, we identify the cybersecurity vendors who published the report. We visit the website of each cybersecurity vendor and collect the CTI-related articles published after $\Delta$. \textbf{Step-3:} Among the collected articles, we only keep those articles where the publisher explicitly mentions the set of MITRE ATT\&CK techniques used in the corresponding cyberattack. In Step-1, we identify $\Delta$ is 18-August-2022. After performing the Step-3, we identify 125\footnote{The list of the URL of these reports can be found at Section 1(E) of supplementary file} unseen CTI reports\footnote{as of the date, 6-March-2023}. On average, we find 12.86 adversarial techniques from each unseen CTI report. The total number of unique techniques across all the unseen CTI reports is 333. We then evaluate the following: (\textbf{EV-A}) to what extent the identified prevalent techniques appear in the set of unseen CTI reports; and (\textbf{EV-B}) to what extent the identified recurring pairs appear in the set of unseen CTI reports.

\textbf{Response to EV-A} We find 18 out of 19 identified prevalent techniques in the unseen CTI reports. The only prevalent technique we did not find is the \textit{T1071.004: DNS}. The mean and median count of prevalent techniques per unseen CTI report is 2.84 and 2, respectively. We identify nine prevalent techniques among the top twenty most-reported techniques among the unseen CTI reports. However, we also find four techniques among the top twenty most reported, which share the same technique id but differs by the sub-technique id. For example, we observe T1204: User Execution is among the top twenty most reported techniques. However, we have T1204.002: Malicious file in the list of 19 prevalent techniques. Thus, if we only consider the technique id, we identify four additional prevalent techniques among the top twenty most reported techniques in the unseen CTI reports. Overall, the evaluation shows that the majority (13 out of 19) of our identified prevalent techniques appear among the top twenty most reported techniques in the set of unseen CTI reports. 


\textbf{Response to EV-B} In RQ2, we identify 425 pairs among 183 unique techniques out of 594 cataloged in ATT\&CK. Collectively, the set of unseen CTI reports mentions 333 ATT\&CK techniques. Among the 425 identified pairs from RQ2, we identify 317 pairs where both the antecedent and consequent techniques are present among the 333 techniques. We refer to the set of 317 pairs as valid pairs. Each of the rest of the 108 pairs, either the antecedent or the consequent technique is not reported by any of the unseen CTI report. Among the 317 valid pairs, we identify 228 of them appears across the unseen CTI reports. We observe at least one valid pair present in 86 unseen CTI reports\footnote{The list of 228 valid pairs can be found at Section 1(G) of supplementary file}. On average, we identify 5.86 valid pairs per unseen CTI report. Overall, the evaluation shows that 54\% of our identified pairs (i.e., 228 out of 425) are found in at least one CTI report, and 69\% of the unseen CTI reports (i.e., 86 out of 125) have at least one recurring pair of techniques. Among the 228 valid pairs identified in the unseen CTI reports, 91 is from \textit{Follow}, 90 from \textit{Same Asset}, 79 from \textit{Implementation Overlap}, 50 from \textit{Happen Together}, 23 from \textit{Require}, 13 from \textit{Platform}, and 9 from \textit{Alternative} relations.

\section{Discussion and takeaways}
\label{sec:discussion}
\textbf{Our study finds the structure of a typical APT attack.} Our study finds that a relatively small number (60) of adversarial techniques are responsible for most (66.2\%) reported adversarial techniques. Especially our study finds 19 prevalent techniques through the analysis of frequency and trend, as reported in Section~\ref{rq1:prevalentTechniques}. The prevalent techniques account for 37.3\% of all the mentions.  Overall we observe the following. First, from a high-level perspective, these small sets of techniques reflect several commonplace behaviors of any APT attacks and malware regardless of the target organizations or malicious goals. Second, adversaries are not always implementing a very complicated and customized malware for their targets. After a successful phishing attempt, the rest of the commonplace malicious work can often be conducted by many openly available remote access tools, C2 frameworks~\cite{S0154, S0363}, credential dumping tools~\cite{S0002, S0349}, modular malware~\cite{S0038}, etc. By combining the first and second observations, organizations can protect themselves by focusing on detecting and mitigating a small set of techniques. MITRE ATT\&CK catalogs 594 techniques. Carefully understanding and prioritizing the defense of all these techniques may incur costs and time for organizations. Instead, our study points out a handful of techniques an average organization should know and prioritize defense around the techniques.  




\textbf{APT attacks misuse system functionalities and essential applications that are hard to be noticed by an user.} We observe four prevalent techniques from TA0002: Execution tactics. One of the techniques is the \textit{T1204.002: Malicious file}, which is eventually performed by a victim user followed by social engineering attempts. Often, the technique is related to the enabling or execution of malicious content (such as a macro) inside a word or spreadsheet document. We recommend organizations apply restrictive policies regarding using or activating executable contents in the document files. We also observe adversaries abuse two system features for persistence: registry and scheduled tasks running in the background. Users are not likely to notice system behavior modification by altering registries and running malicious programs through scheduled tasks.  We recommend organizations apply restrictive policies regarding any modification performed at the registries or creation of scheduled tasks.  We also observe adversaries abusing several essential applications, usually found in desktop operating systems, primarily office applications, browsers, remote work and communication tools, and cloud storage applications. We observe phishing techniques are usually associated with hidden malicious executables in text, spreadsheet, image, and PDF documents. We also observe adversaries turn internet browsers into a server to maintain C2 communication. Dumped credentials are often abused to log in to remote access protocols (e.g., RDP, VNC, SSH, VPN) and software (e.g., Windows Remote Desktop). Locally installed version control systems (e.g., Github) and cloud storage synchronization applications (e.g., Dropbox, Onedrive) often transfer local files from victim systems to adversary-controlled locations. Day-to-day use of these applications and being run in the background often make misuse of these applications difficult to notice, even by anti-malware systems. We speculate that APT attacks target personal computers more than server-based applications. Their target is primarily not the applications running on the servers but the human users doing their day-to-day office work. 


\textbf{Implication for IT organizations:} CTI reports usually cover high-profile cyberattacks, such as attacks on the government, utility sector, and fin-tech organizations. The reports, however, do not specifically document any lack of mitigation, countermeasures, or security control failures related to attacks. As the attacks are performed on high-profile organizations, we can assume that the victim organizations must have deployed security countermeasures. However, average APT attacks not being sophisticated makes us speculate that: (a) either the organizations, overall, did not deploy sufficient extent of security countermeasures, or (b) they acquired security products and applied countermeasures but failed to put together an adequate level of defense in practice. 

\textbf{APT attacks need to be detected in the early stages}. We identify five prevalent techniques from the \textit{TA0007: Discovery} tactic. As per MITRE ATT\&CK, techniques from the tactic do not have any specific mitigation – because these techniques include running commands which only require read permission and minimal privileges. We recommend practitioners deploy mechanisms for detecting the execution of these commands, although they might look benign. We observe secondary stage malware is downloaded based on the obtained information from the techniques from the \textit{TA0007: Discovery} tactic. The observation suggests that detecting primary-stage malware, which collects system-specific information, is critical for blocking APT attacks in their early stages. 

\textbf{Early-stage prevention of APT attacks increasingly depends on training users about cybersecurity best practices.} We observe that social engineering vectors, such as phishing attachments and links, are the primary ways for adversaries to make an initial compromise. Thus, computer users in organizations may act as one of the weakest links toward defense. Although modern browsers, email service providers, and anti-malware software can help users to detect suspicious websites and phishing attempts. We urge organizations to build awareness of cybersecurity best practices among users. 

\textbf{Implication for cyberthreat hypothesis hunting}. The identified recurring pairs of adversarial techniques can aid practitioners in cyberthreat hunting. As APT attacks take several months to be detected, practitioners can utilize the identified pairs to find any ongoing adversarial attempt. For example, the execution of \verb|systeminfo| command may not trigger any alert because the command does not execute any code or modify any system settings. However, from the identified pairs, we can hypothesize that if the command is executed simultaneously with other TA0007: Discovery-related commands, such as \verb|ipconfig|, \verb|ps|, etc., an adversary will probably execute the commands. The hypothesis can aid practitioners in triangulating and triaging the malicious process executing the behaviors. The recurring pairs can also aid practitioners in emulating adversary behaviors for performing red vs. blue team exercises. Practitioners can also check the effectiveness of their security products and user training programs against the prevalent techniques. 


\textbf{Detection and mitigation strategies should be adaptive to accommodate the change in the threat landscape.} Our study identifies recurring pairs of techniques documented by MITRE ATT\&CK. However, the threat landscape changes and techniques used by adversaries also depend on the adversaries' expertise and the weakness of the victim environment. Hence, organizations can run co-occurrence analyses on their environment and adapt detection and mitigation strategies that suit organizations' workflow, system architecture, and security enforcement. Organizations can also conduct longitudinal analysis to discover the correlation between adversarial techniques and the limitations of security solutions.


\section{Threats to validity}
\label{sec:threats}
We discuss several limitations of the study in this section. The dataset we use reflects the adversarial techniques documented by MITRE ATT\&CK. The dataset contains only the techniques identified by security experts and then reported after cyberattacks. Consequently, adversaries may have used other techniques that were not detected and remain unreported. Thus, the dataset only reflects a subset of the techniques adversaries used. MITRE ATT\&CK mapped the techniques from publicly-reported documents using an automated manner~\cite{mitre-tram}, which may have introduced mapping bias in the dataset. Nonetheless, the study approximates how cybercrime groups and malware are leveraging techniques on various types of organizations across the globe. Practitioners and defenders can use the methodology and findings of the paper as a starting point to investigate adversarial techniques in their environment, derive mitigation strategies, and improve security practices. Hence, the approach presented in the study requires further data collection and ground truth construction from numerous organizations across various parts of the globe. The coordinated effort among organizations and cybersecurity vendors can capture generalized information on how techniques interact. 

\section{Conclusion}
\label{sec:conclusion}

This study aims to understand how attackers use adversarial techniques through malware in APT attacks. To this end, we collect the technical and forensic analysis of past APT attacks described in 667 CTI reports. We systematically investigate 594 MITRE ATT\&CK techniques reported in these 667 CTI reports. We analyzed the frequency and trend of adversarial techniques, followed by a qualitative analysis of the implementation of techniques. We perform association rule mining to identify pairs of techniques appearing in CTI reports. We then perform qualitative analysis to identify the underlying relations among the techniques in recurring pairs. We identify (a) a set of 19 prevalent techniques observed in APT attacks, (b) a set of 425 recurring pairs among adversarial techniques, and (c) seven types of relations among the adversarial techniques. We make our dataset and analysis scripts available for researchers to rerun the analysis on future versions of ATT\&CK and different datasets. Organizations can use our approach as a starting point to formulate proactive defensive strategies. We also advocate researchers and practitioners make synchronized efforts to collect more data to draw a comprehensive picture of how cyberattack happens and how we can make organizations more secure.

\begin{acks}
This work is partly supported by the NSA Science of Security award H98230-17-D-0080. Any findings and opinions expressed in this material are those of the authors and do not necessarily reflect the views of the funding agencies.
\end{acks}

\bibliographystyle{ACM-Reference-Format}
\bibliography{main}



\end{document}